\input harvmac
\noblackbox


\font\ticp=cmcsc10  
\def\Title#1#2{\rightline{#1}\ifx\answ\bigans\nopagenumbers\pageno0\vskip1in
\else\pageno1\vskip.8in\fi \centerline{\titlefont #2}\vskip .5in}
 
scaled\magstep3

scaled\magstep3 
 
scaled\magstep3 
 
scaled\magstep3 \font\ticp=cmcsc10 \font\ttsmall=cmtt10 at 8pt

\input epsf
\ifx\epsfbox\UnDeFiNeD\message{(NO epsf.tex, FIGURES WILL BE
IGNORED)}
\def\figin#1{\vskip2in}
\else\message{(FIGURES WILL BE INCLUDED)}\def\figin#1{#1}\fi
\def\ifig#1#2#3{\xdef#1{Fig.\the\figno}
\goodbreak\topinsert\figin{\centerline{#3}}%
\smallskip\centerline{\vbox{\baselineskip12pt
\advance\hsize by -1truein\noindent{\bf Fig.~\the\figno:} #2}}
\bigskip\endinsert\global\advance\figno by1}

\input labeldefs.tmp
\writedefs

%
%
\def\la{\mathrel{\mathpalette\fun <}}
\def\ga{\mathrel{\mathpalette\fun >}}
\def\fun#1#2{\lower3.6pt\vbox{\baselineskip0pt\lineskip.9pt
  \ialign{$\mathsurround=0pt#1\hfil##\hfil$\crcr#2\crcr\sim\crcr}}}
\relax

\def\[{\left [}
\def\]{\right ]}
\def\({\left (}
\def\){\right )}
\def\eg{{\it e.g.}}
\def\ie{{\it i.e.}}
\def\cf{{\it cf.}}
\def\etc{{\it etc.}}
\def\om{\omega}
\def\Om{\Omega}
\def\p{\partial}

\def\pd{{\dot \phi}}
\def\pdd{{\ddot \phi}}
\def\ad{{\dot a}}
\def\add{{\ddot a}}
\def\psd{{\dot \psi}}
\def\psdd{{\ddot \psi}}
\def\bd{{\dot b}}
\def\bdd{{\ddot b}}

\def\p0{\phi_0}
\def\po{\phi_0}
\def\p1{\phi_1}
\def\e{\varepsilon}
\def\l{\ell}

\def\a{\alpha}

\def\CO{{\cal O}}

\def\CN{{\cal N}}
\def\CR{{\cal R}}
\def\S{{\bf S}}
\def\R{{\bf R}}
\def\A5S5{AdS$_5 \times  \S^5$}
\def\Tco{T_{\rm trust}}
\def\Rco{R_{\rm trust}}
\def\Mbh{M_{\rm BH}}
\def\Mc{M_{\rm config}}
\def\mBF{m_{\rm BF}}
\def\half{{1 \over 2}}
\def\rs{R_s}
\def\rp{r_+}
\def\rmx{r_{h2}}
\def\Rp{r_{h3}}
\def\rpa{r_{h0}}
\def\rpt{r_{h1}}
\def\rsz{R_{s0}}
\def\rso{R_{s1}}

\def\ss{\sigma_s}
\def\ps{\phi_s}
\def\ep{\epsilon}
\def\ie{{\it i.e.}}


\lref\penr{ R.~Penrose, {\it Gravitational Collapse: The Role Of
General Relativity}, Riv.\ Nuovo Cim.\  {\bf 1}, 252 (1969) [Gen.\
Rel.\ Grav.\  {\bf 34}, 1141 (2002)].}

\lref\hhm{ T.~Hertog, G.~T.~Horowitz and K.~Maeda, {\it Negative
energy in string theory and cosmic censorship violation},
[arXiv:hep-th/0310054].}

\lref\hhmgr{ T.~Hertog, G.~T.~Horowitz and K.~Maeda, {\it Generic
cosmic censorship violation in anti de Sitter space},
[arXiv:gr-qc/0307102].}

\lref\kraus{ M.~Gutperle and P.~Kraus, {\it Numerical Study of
Cosmic Censorship in String Theory}, arXiv:hep-th/0402109}

\lref\gonzalez{ M.~Alcubierre, J.~A.~Gonzalez, M.~Salgado and
D.~Sudarsky, {\it The cosmic censor conjecture: Is it generically
violated?}, arXiv:gr-qc/0402045.}

\lref\he{ S.~W.~Hawking and G.~F~.R.~Ellis, {\it The Large Scale
Structure of Space-Time}, Cambridge University Press, (1973).}

\lref\wald{ R.~M.~Wald, {\it Gravitational collapse and cosmic
censorship}, [arXiv:gr-qc/9710068].}

\lref\choptuik{ M.~W.~Choptuik,
 {\it Universality And Scaling In Gravitational Collapse Of A Massless
Scalar
 Field},
Phys.\ Rev.\ Lett.\  {\bf 70}, 9 (1993).}

\lref\joshi{ P.~S.~Joshi and I.~H.~Dwivedi,
 {\it Naked Singularities In Spherically Symmetric Inhomogeneous
Tolman-Bondi Dust Cloud Collapse}, Phys.\ Rev.\ D {\bf 47}, 5357
(1993) [arXiv:gr-qc/9303037].}

\lref\maeda{ K.~Maeda, T.~Torii and M.~Narita,
 {\it Do naked singularities generically occur in generalized theories
of
 gravity?},
Phys.\ Rev.\ Lett.\  {\bf 81}, 5270 (1998) [arXiv:gr-qc/9810081].}

\lref\juan{ J. Maldacena, {\it The Large N Limit of Superconformal
Field Theories and Supergravity}, Adv. Theor. Math. Phys. {\bf 2}
(1998) 231, [arXiv:hep-th/9711200].}

\lref\magoo{ O. Aharony, S.S. Gubser, J. Maldacena, H. Ooguri, Y.
Oz, {\it Large N Field Theories, String Theory and Gravity}, Phys.
Rept. {\bf 323} (2000) 183, [arXiv:hep-th/9905111].}

\lref\witten{ E. Witten, {\it Anti De Sitter Space And
Holography}, Adv. Theor. Math. Phys. {\bf 2} (1998) 253,
[arXiv:hep-th/9802150].}

\lref\gkp{ S. Gubser, I. Klebanov, and A. Polyakov, {\it Gauge
Theory Correlators from Non-Critical String Theory}, Phys. Lett.
{\bf B428} (1998) 105, [arXiv:hep-th/9802109].}

\lref\torii{ T.~Torii, K.~Maeda and M.~Narita, {\it Scalar hair on
the black hole in asymptotically anti-de Sitter spacetime}, Phys.\
Rev.\ D {\bf 64}, 044007 (2001).}

\lref\breit{P.~Breitenlohner and D.~Z.~Freedman, {\it Stability In
Gauged Extended Supergravity}, Annals Phys.\  {\bf 144}, 249
(1982);
{\it Positive Energy In Anti-De Sitter Backgrounds And Gauged
Extended Supergravity}, Phys.\ Lett.\ B {\bf 115}, 197 (1982).}

\lref\mezin{ L.~Mezincescu and P.~K.~Townsend, {\it Stability At A
Local Maximum In Higher Dimensional Anti-De Sitter Space And
Applications To Supergravity}, Annals Phys.\  {\bf 160}, 406
(1985).}

\lref\fhks{ L.~Fidkowski, V.~Hubeny, M.~Kleban and S.~Shenker,
{\it The black hole singularity in AdS/CFT},
[arXiv:hep-th/0306170].}

\lref\sw{ N.~Seiberg and E.~Witten, {\it The D1/D5 system and
singular CFT}, JHEP {\bf 9904}, 017 (1999) [arXiv:hep-th/9903224].
}

\lref\mm{ J.~Maldacena and L.~Maoz, {\it Wormholes in AdS},
[arXiv:hep-th/0401024].
}

\lref\ms{ J.~M.~Maldacena and A.~Strominger, {\it AdS(3) black
holes and a stringy exclusion principle}, JHEP {\bf 9812}, 005
(1998) [arXiv:hep-th/9804085].
}

\lref\dkss{ S.~Deger, A.~Kaya, E.~Sezgin and P.~Sundell, {\it
Spectrum of D = 6, N = 4b supergravity on AdS(3) x S(3)}, Nucl.\
Phys.\ B {\bf 536}, 110 (1998) [arXiv:hep-th/9804166].
}

\lref\nics{ H.~Nicolai and H.~Samtleben, {\it Kaluza-Klein
supergravity on AdS(3) x S(3)}, JHEP {\bf 0309}, 036 (2003)
[arXiv:hep-th/0306202].
}

\lref\bce{ J.~Babington, D.~E.~Crooks and N.~Evans, {\it A
non-supersymmetric deformation of the AdS/CFT correspondence},
JHEP {\bf 0302}, 024 (2003) [arXiv:hep-th/0207076].
}

\lref\clpst{ M.~Cvetic, H.~Lu, C.~N.~Pope, A.~Sadrzadeh and
T.~A.~Tran, {\it Consistent SO(6) reduction of type IIB
supergravity on S(5)}, Nucl.\ Phys.\ B {\bf 586}, 275 (2000)
[arXiv:hep-th/0003103].
}

\lref\dfghm{ O.~DeWolfe, D.~Z.~Freedman, S.~S.~Gubser,
G.~T.~Horowitz and I.~Mitra, {\it Stability of AdS(p) x M(q)
compactifications without supersymmetry}, Phys.\ Rev.\ D {\bf 65},
064033 (2002) [arXiv:hep-th/0105047].
}

\lref\krn{ H.~J.~Kim, L.~J.~Romans and P.~van Nieuwenhuizen, {\it
The Mass Spectrum Of Chiral N=2 D = 10 Supergravity On S**5},
Phys.\ Rev.\ D {\bf 32}, 389 (1985).
}

\lref\dijk{ R.~Dijkgraaf, {\it Instanton strings and hyperKaehler
geometry}, Nucl.\ Phys.\ B {\bf 543}, 545 (1999)
[arXiv:hep-th/9810210].
}

\lref\dmwy{ A.~Dhar, G.~Mandal, S.~R.~Wadia and K.~P.~Yogendran,
 {\it D1/D5 system with B-field, noncommutative geometry and the
CFT of the  Higgs branch}, Nucl.\ Phys.\ B {\bf 575}, 177 (2000)
[arXiv:hep-th/9910194].
}

\lref\klebw{ I.~R.~Klebanov and E.~Witten, {\it AdS/CFT
correspondence and symmetry breaking}, Nucl.\ Phys.\ B {\bf 556},
89 (1999) [arXiv:hep-th/9905104].
}

\lref\fpw{ D.~Z.~Freedman, S.~S.~Gubser, K.~Pilch and
N.~P.~Warner, {\it Continuous distributions of D3-branes and
gauged supergravity}, JHEP {\bf 0007}, 038 (2000)
[arXiv:hep-th/9906194].
}

\lref\qnm{ G.~T.~Horowitz and V.~E.~Hubeny,
 {\it Quasinormal modes of AdS black holes and the approach to thermal
equilibrium}, Phys.\ Rev.\ D {\bf 62}, 024027 (2000)
[arXiv:hep-th/9909056].
}

\lref\overcharge{ V.~E.~Hubeny, {\it Overcharging a Black Hole and
Cosmic Censorship}, Phys.\ Rev.\ D {\bf 59}, 064013 (1999)
[arXiv:gr-qc/9808043].
}

\lref\PolchinskiUF{ J.~Polchinski and M.~J.~Strassler, ``The
string dual of a confining four-dimensional gauge theory,''
arXiv:hep-th/0003136.
}

\lref\HorowitzNW{ G.~T.~Horowitz and J.~Polchinski, ``A
correspondence principle for black holes and strings,'' Phys.\
Rev.\ D {\bf 55}, 6189 (1997) [arXiv:hep-th/9612146].
}

\lref\SusskindWS{ L.~Susskind, ``Some Speculations About Black
Hole Entropy In String Theory,'' arXiv:hep-th/9309145.
}

\lref\DamourWM{ T.~Damour and M.~Henneaux, ``Chaos in superstring
cosmology,'' Phys.\ Rev.\ Lett.\  {\bf 85}, 920 (2000)
[arXiv:hep-th/0003139].
}

\lref\JohnsonQT{ C.~V.~Johnson, A.~W.~Peet and J.~Polchinski, {\it
Gauge theory and the excision of repulson singularities}, Phys.\
Rev.\ D {\bf 61}, 086001 (2000) [arXiv:hep-th/9911161].
}

\lref\garfinkle{D.~Garfinkle, {\it Numerical simulation of a
possible counterexample to cosmic censorship},
[arXiv:gr-qc/0403078]. }

\lref\frolov{A.~Frolov, {\it in preparation.}}

\lref\hhmrev{ T.~Hertog, G.~T.~Horowitz and K.~Maeda, {\it Update
on Cosmic Censorship Violation in AdS}, [arXiv:gr-qc/0405050].}

\lref\dafermos{ M.~Dafermos, {\it A note on naked singularities
and the collapse of self-gravitating Higgs fields},
[arXiv:gr-qc/0403033].}


%
\baselineskip 16pt \Title{\vbox{\baselineskip12pt \line{\hfil
SU-ITP-04-07} \line{\hfil UCB-PTH-04/05} \line{\hfil LBNL-54620}
\line{\hfil \tt hep-th/0403198} }} {\vbox{ {\centerline{Comments
on Cosmic Censorship in AdS/CFT} }}} \centerline{\ticp Veronika E.
Hubeny$^a$, Xiao Liu$^{a,b}$,
 Mukund Rangamani$^{c,d}$,
 and Stephen Shenker$^a$
\footnote{}{\ttsmall
 veronika@itp.stanford.edu, liuxiao@itp.stanford.edu,}
\footnote{}{\ttsmall mukund@socrates.berkeley.edu,
sshenker@stanford.edu}}
\bigskip
\centerline {\it $^a$ Department of Physics, Stanford University,
Stanford, CA 94305, USA}\centerline{\it $^b$ SLAC, Stanford
University, Stanford, CA 94309, USA} \centerline{\it $^c$
Department of Physics, University of California, Berkeley, CA
94720, USA} \centerline{\it $^d$ Theoretical Physics Group, LBNL,
Berkeley, CA 94720, USA}

\bigskip
\centerline{\bf Abstract}
\bigskip

Recently Hertog, Horowitz, and Maeda (HHM) (hep-th/0310054) have
proposed that cosmic censorship can be violated in the AdS/CFT
context.   They  argue that for certain initial data there is
insufficient energy available to make a black hole whose horizon
is big enough to cloak the singularity that forms.   We have
investigated this proposal in the models HHM discuss and have thus
far been unable to find initial data that provably satisfy this
criterion, despite our development of an improved lower bound on
the size of the singular region.  This is consistent with recent
numerical results (hep-th/0402109). For certain initial data, the
energies of our configurations are not far above the lower bound
on the requisite black hole mass, and so it is possible that in
the exact time development naked singularities do form. We go on
to argue that the finite radius cut-off AdS$_5$ situation
discussed by HHM displays instabilities when the full 10D theory
is considered. We propose an AdS$_3$ example that may well be free
of this instability.

\smallskip

\Date{March 2004}

\newsec{Introduction}

Curvature singularities are a rather generic feature in general
relativity. As demonstrated by the singularity theorems \he, even
relatively mild generic initial data can evolve to a singularity.
Indeed, in nature, we expect singularities inside black holes as
well as cosmological ones. Of course, the laws of general
relativity break down at a singularity, thereby precluding us from
evolving the classical spacetime into its causal future. The
feeling that the collapse to a singularity in one region of
spacetime should not terminate the classical evolution in regions
arbitrarily far away prompted Penrose \penr\ to conjecture that
these singularities are generically cloaked by event horizons, and
therefore hidden inside black holes. In other words, there is some
{\it cosmic censor}, which prevents the formation of {\it naked
singularities} (\ie, those visible to distant observers). More
specifically, the Cosmic Censorship Conjecture states that smooth
and generic initial data with ``reasonable'' matter cannot evolve
to a naked singularity\foot{ In fact there are two versions: the
{\it weak} cosmic censorship conjecture precludes the
singularities to be visible asymptotically; while the {\it strong}
cosmic censorship conjecture disallows formation of timelike
singularities, even if they are inside black holes.  Here we shall
consider only the weak version.}.

The Cosmic Censorship Conjecture, first formulated over three
decades ago, has enjoyed a vibrant history (for a recent review of
the subject, see \eg\ \wald ) and remains one of the most
important and intriguing questions in general relativity today.
While no proof has yet been found, cosmic censorship is often
assumed to hold. On the other hand, numerous ``counter-examples''
have been proposed; typically this has led to tighter formulations
of the conjecture. For example, naked singularities can be
produced with fine-tuned initial conditions \choptuik, with
pressureless matter \joshi, \etc, whereas various proposals with
generic and reasonable initial conditions, such as overcharging a
black hole \overcharge, are too delicate to prove the production
of naked singularities. Recently Hertog, Horowitz, and Maeda
\hhmgr\ have argued that cosmic censorship can be violated for
configurations with a scalar field coupled to gravity in
asymptotically AdS spacetimes. This would give one of the first
examples of cosmic censorship violation using generic initial
conditions with reasonable matter\foot{ See \maeda\ for cosmic
censorship violation in an asymptotically dS case.}.

The AdS boundary conditions facilitate formation of naked
singularities, as we will review below.  More intriguingly,
however, they are of importance in string theory: According to the
well-known AdS/CFT correspondence \refs{\juan, \magoo, \witten,
\gkp}, Type IIB string theory on asymptotically \A5S5\ spacetime
is dual to a 4-dimensional $\CN = 4$ Super Yang-Mills gauge theory
``living on the boundary" of AdS. Since the correspondence holds
at the quantum level, we expect that any naked singularity
produced in the bulk of AdS will be resolved by the gauge theory.
After all, the CFT evolution is well-defined and should not
terminate at some finite time when the gauge theory comes into
causal contact with the singularity. So embedding naked
singularities in AdS/CFT would provide a robust way to study
quantum gravity in these extreme environments.  The current
methods \fhks\ for studying spacelike singularities in AdS/CFT
focus on the black hole singularity which lives behind a horizon,
and so requires delicate techniques to investigate.

As a first step to the program of using the dual field theory
description to see how singularities are resolved in a quantum
theory of gravity, one needs to construct simple configurations in
string theory which evolve to a naked singularity.

Unfortunately, the results of \hhmgr\ cannot be directly applied
to this endeavor, because the action used there is not one derived
from string theory.  A suitable potential for the scalar field is
chosen by hand rather than deriving it from the appropriate
supergravity action. However, in a very interesting follow-up
paper \hhm\ , Hertog, Horowitz, and Maeda (HHM) do examine cosmic
censorship in the AdS/CFT context.  In particular, they argue that
naked singularities will be produced generically even in 5D $\CN =
8$ supergravity, a consistent truncation of the low energy limit
of string theory on AdS$_5$ $\times \S^5$.  The essence of their
argument is to produce a simple configuration that reliably
evolves into a singularity, and then show that there is
insufficient energy available to make a black hole whose horizon
is big enough to cloak the singularity. Recently, this claim has
been challenged in \kraus, whose numerical study of the HHM
process suggests\foot{We should point out, though, that this
numerical study only indicates the presence of trapped surfaces.
While this is suggestive, a compelling demonstration might well
require evolving the initial data far enough into the future to
see the exterior metric settle into the Schwarzschild-AdS black
hole geometry.} formation of black holes rather than naked
singularities.

In this note, we study the parameter space for initial data with
the purpose of identifying a region where one can reliably argue
that naked singularities  form and moreover the dual CFT is
well-behaved. In the next section, we give a brief review of the
two HHM papers. In Section 3 we discuss the infinite radius non
cut-off AdS/CFT  context, consistently truncated to 5D
supergravity, studied by HHM.  We are unable to find any initial
data where we can rigorously show that the configuration possesses
too little energy to make a black hole of the requisite size. The
energy excess is small in certain parameter regimes and so the
possibility exists that a more accurate (and hence larger)
estimate of the size of the singular region will show that a naked
singularity must form.

In Section 4 we turn to the finite radius cut-off AdS version
introduced in \hhm (v2).   The situation is similar to the
infinite radius situation. In addition, here we show that the full
10D theory is unstable to D3-branes moving to the boundary, using
both bulk and boundary CFT arguments.  The true late time behavior
of the theory will be dominated by this instability and so
predictions for the full theory from the five dimensional analysis
are problematic.

 In Section 5, we focus on a special case of cut-off
3-dimensional AdS, where the instability seems absent. There may
well be other strong 10 dimensional dynamics present in this
system though.  Again we do not find any initial data that
reliably produce naked singularities.

We end with a summary and some discussion in Section 6. Most of
the detailed calculations and numerical estimates are relegated to
Appendices, where among other things we refine and extend the
estimates of HHM to a more general set-up in $d$ dimensions.

\newsec{Review of HHM}

In \hhmgr, Hertog, Horowitz, and Maeda consider 4-dimensional
gravity coupled to a scalar field $\phi$ with the potential
$V(\phi)$.  The potential, sketched in Fig.1a, has two negative
minima, a global one at $\phi=0$ and a local one at $\phi_1$; and
the solutions are chosen so as to asymptotically approach $\phi_1$
(\ie, $V(\phi_1)$ determines the AdS radius of the resulting
asymptotically AdS spacetime).  As a physical requirement, the
potential $V(\phi)$ is chosen so as to satisfy the positive energy
theorem for all solutions with these boundary conditions. More
specifically, the (time-symmetric and spherically symmetric)
initial conditions consist of the field profile as sketched in
Fig.1b: the field is near the global minimum $\phi \approx 0$ in
some interior region $r \le R_0$, and then smoothly asymptotes to
the local minimum $\phi_1$.

\ifig\figHHMgr{Potential $V(\phi)$ and the field profile $\phi(r)$
used by HHM  \hhmgr\ in the purely gravitational context.}
{\epsfxsize=12cm \epsfysize=3.7cm \epsfbox{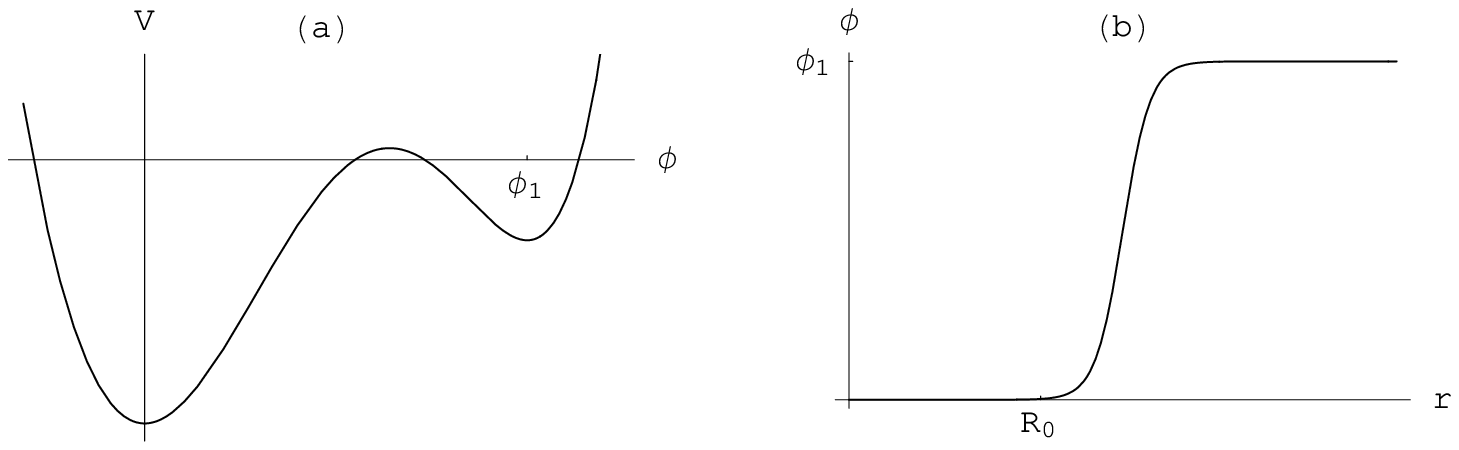}}

In the central region, where the field is homogeneous and
isotropic, the evolution is determined by the standard FRW
equations.  In particular, since a homogeneous scalar field
rolling down a potential to a negative minimum becomes singular,
our spacetime will develop a spacelike singularity in the domain
of dependence of the central region.  Will this singularity be
cloaked by an event horizon, or can it be naked?  In order for the
singularity to be hidden inside a black hole, the black hole
horizon has to be large enough to encompass the singularity, so
correspondingly the black hole mass has to be large enough. On the
other hand, by mass conservation, if the initial configuration
forms a black hole, the black hole mass can be no larger than the
initial mass. Thus, a naked singularity must result if the mass
$\Mc$ of the initial configuration is smaller than the mass $\Mbh$
of the smallest black hole large enough to cloak the ensuing
singularity.

By cleverly tuning the potential, the authors of \hhmgr\ show that
there exists an open set of initial conditions which satisfy this
criterion. In particular, by making the interior region very
large, $R_0 \gg 1$, the black hole size must correspondingly scale
as $R_0^3$. Although generically the mass of the initial
configuration also scales as $R_0^3$, by tuning the potential,
they are able to cancel this leading contribution, so that instead
the mass scales only as $R_0$.  Hence, by taking $R_0$ large
enough, it appears that they can easily satisfy the naked
singularity condition, $\Mc < \Mbh$.\foot{The fact that $\Mbh \sim
r_+^3$ for large horizon radii $r_+$ is due to the AdS boundary
conditions; to form a black hole of a given size, more mass is
required in AdS than in flat space. Since in the latter
 $\Mbh \sim r_+$, more detailed study is
necessary to certify whether a similar mechanism would allow the
formation of a naked singularity in an asymptotically flat
spacetime.}
This argument is not complete however, as we will explain in
Appendix B.6.  This same issue is also discussed in \hhmrev .
Evidence pointing toward this conclusion was presented in
\garfinkle . \foot{Recently the authors of \gonzalez\ have argued
that there is a loophole in this argument: the singularity need
not be cloaked by a black hole which asymptotically looks large
and static, but instead the field may act as a forever expanding
domain wall of negative energy, leaving room inside for a large
black hole. This can't happen in a system satisfying the positive
energy theorem, so we do not consider it further. We thank Gary
Horowitz for pointing this out to us; \cf \hhmrev . }

To consider the analogous situation in the AdS/CFT context, one no
longer has the freedom to pick an arbitrary potential, which makes
the arguments for cosmic censorship violation much more involved.
Nevertheless, HHM \hhm\ argue that naked singularities can still
be formed generically. This relies on constructing configurations
which, as in the previous case, collapse to a singularity, but
from the bulk point of view initially have negative energy, and
therefore seemingly cannot form a black hole.

In particular, HHM use the well-known fact that in AdS, scalar
fields can have negative mass squared, but nevertheless can be
stable, provided they satisfy the Breitenlohner-Freedman (BF)
bound $m^2 \ge \mBF^2 = -{(d-1)^2 \over 4}$ in AdS$_d$
\refs{\breit,\mezin}. While the positive energy theorem is
necessarily satisfied for $m^2 > \mBF^2$, for fields saturating
the BF bound $m^2 = \mBF^2$, the total bulk (ADM) energy (which is
normalized such that pure AdS has zero energy) can be arbitrarily
negative \hhm.  In the bulk, this ``bulk energy''
 is what governs the evolution in the domain of
dependence of the initial data surface\foot{More specifically, as
explained in Appendix A, if we write the spatial metric on the
initial slice as in \sigmamet, $ds^2_{\Sigma} = \(1 - {m(r)\over
3\, \pi^2 \, r^2 } + r^2 \)^{-1} dr^2 + r^2  \, d\Omega_3^2$, then
the ADM mass is given simply by $\lim_{r \to \infty} m(r)$; and
the bulk evolution is sensitive to the local metric, \ie, to the
function $m(r)$. The generalization to $d$ dimensions can be found
in \sigmametd.}. However, the dual CFT energy corresponds to the
bulk energy {\it plus} a finite counter-term, which may be viewed
as the boundary contribution. This boundary term vanishes when the
field falls off faster than the fall-off given by the BF bound,
but it is present for the BF-saturating fields. This ensures that
the CFT energy is positive and conserved.

HHM concentrate on the $\CN = 8$ gauged supergravity on AdS$_5$,
which arises as the low energy limit of string theory with AdS$_5
\times \S^5$ boundary conditions, and is conjectured to be a
consistent 5 dimensional truncation of 10 dimensional Type IIB
supergravity \clpst.

In  global coordinates, with AdS$_5$ given by
\eqn\AdS { ds^2 = -(1+r^2) \, dt^2 + {dr^2 \over 1+r^2} + r^2 \,
d\Om_3^2  \ ,}
the solutions to the scalar wave equation $(\nabla^2 - m^2) \,
\phi = 0$ fall off asymptotically as $\phi(r) \sim
r^{1/\lambda_{\pm}}$ where $\lambda_{\pm} \equiv 2 \pm
\sqrt{4+m^2}$.  When $m^2 = \mBF^2 = -4$, the solutions behave as
$1/r^2$ and $\log r /r^2$. Since the latter is non-normalizable
(unless one considers the cut-off version of the correspondence as
discussed in Section 4), HHM focus on the former. From the higher
dimensional point of view, the BF-saturating scalars correspond to
the $\l = 2$ modes on the $\S^5$ \krn. There is one special field
which does not source any of the others, so HHM set all other
fields to zero and concentrate on this single scalar field,
$\phi$.  The action (with AdS radius set to 1) then reduces to
\eqn\HHMaction{ S = \int \sqrt{-g} \[ \half R - \half (\nabla
\phi)^2 - V(\phi) \] \ ,}
where the potential $V(\phi)$, plotted in Fig.2a, is given by
\eqn\HHMpot{ V(\phi) = -2 \, e^{2 \phi / \sqrt{3}} - 4 \, e^{-
\phi / \sqrt{3}} \ . }
\ifig\figHHM{Supergravity potential $V(\phi)$ and the field
profile $\phi(r)$ used by HHM  \hhm\ in the AdS/CFT context.}
{\epsfxsize=12cm \epsfysize=3.7cm \epsfbox{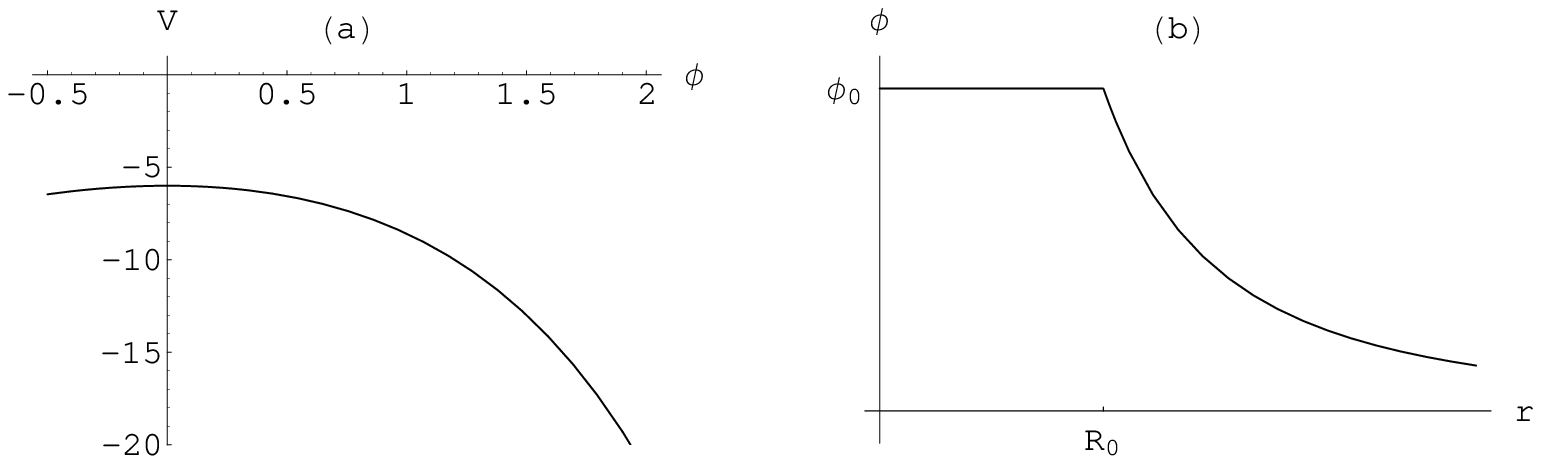}}

The particular field configuration that HHM consider as initial
conditions (which they again take to be time-symmetric and
spherically symmetric) is plotted in Fig.2b, and is given by

\eqn\HHMphi{ \phi(r)  = \po \equiv { A \over R_0^{2}}  \qquad r
\le R_0 \ , \qquad \;\;\;\; \phi(r)= {A \over r^{2}} \qquad r >
R_0 \ . }
In the domain of dependence of the internal region $r \le R_0$ the
field evolves according to the FRW equations to a spacelike
singularity. The detailed evolution is described in Appendix B for
$d$ dimensions; the HHM set-up corresponds to $d=5$. Physically,
the field in the central region rolls down the potential and
diverges. However, since the asymptotic field is pinned to the top
of the potential, the spacetime is asymptotically AdS.

The main question concerns the evolution of the field outside the
domain of dependence of the internal region.  There are several
a-priori possibilities as to what can happen:  1) The spacelike
singularity joins on to the spacelike black hole singularity, \ie,
the field configuration collapses to a black hole, and the CFT
doesn't ``see" the singularity at any finite time\foot{ There may
of course still be signatures of the black hole singularity, as
discussed in \fhks, but these do not correspond to local
observables on the boundary.}. 2) The singularity stops inside the
bulk or becomes timelike (these are indistinguishable from the
point of view of the classical evolution, since the evolution in
the future domain of influence of the singularity is not
well-defined).  The CFT then ``sees" this naked singularity at
some finite time (presumably, there are local observables which
reflect the causal contact with the high curvature region), but
since the evolution in the CFT is well defined, ultimately this
should ``resolve" the naked singularity. This is the case that HHM
argue for. 3) The singularity continues and eventually ``hits" the
boundary at a finite CFT time--a Big Crunch.  This is the most
drastic (and therefore presumably least likely) outcome, for then
the CFT would seem to end or change abruptly. Remarkably, Dafermos
\dafermos\ has proven that in the case of scalar fields with
potentials that are bounded from below (as in the example of
\hhmgr) if cosmic censorship is violated, then a Big Crunch must
necessarily occur.

\newsec{AdS/CFT without a cut-off}

To prove the formation of naked singularities, HHM first argued
that the initial configuration (which is guaranteed to evolve to a
singularity) does not have sufficient energy to form a black hole.
In particular, the ADM energy of the initial configuration is
negative.  And although the CFT energy is guaranteed to be
positive, it is the bulk energy (given by the bulk metric) which
determines the bulk evolution. Unfortunately, this simple
kinematic argument is not easily applicable for BF-saturating
fields and  HHM have already pointed out a possible loophole in
this argument \hhm (v2). Essentially, as we explain below, the
bulk energy rises too fast, and the conserved energy is too large.

\ifig\pdns{Causal diagram of the formation of the singularity for
small initial fields $\po$.} {\epsfxsize=6.8cm \epsfysize=5.1cm
\epsfbox{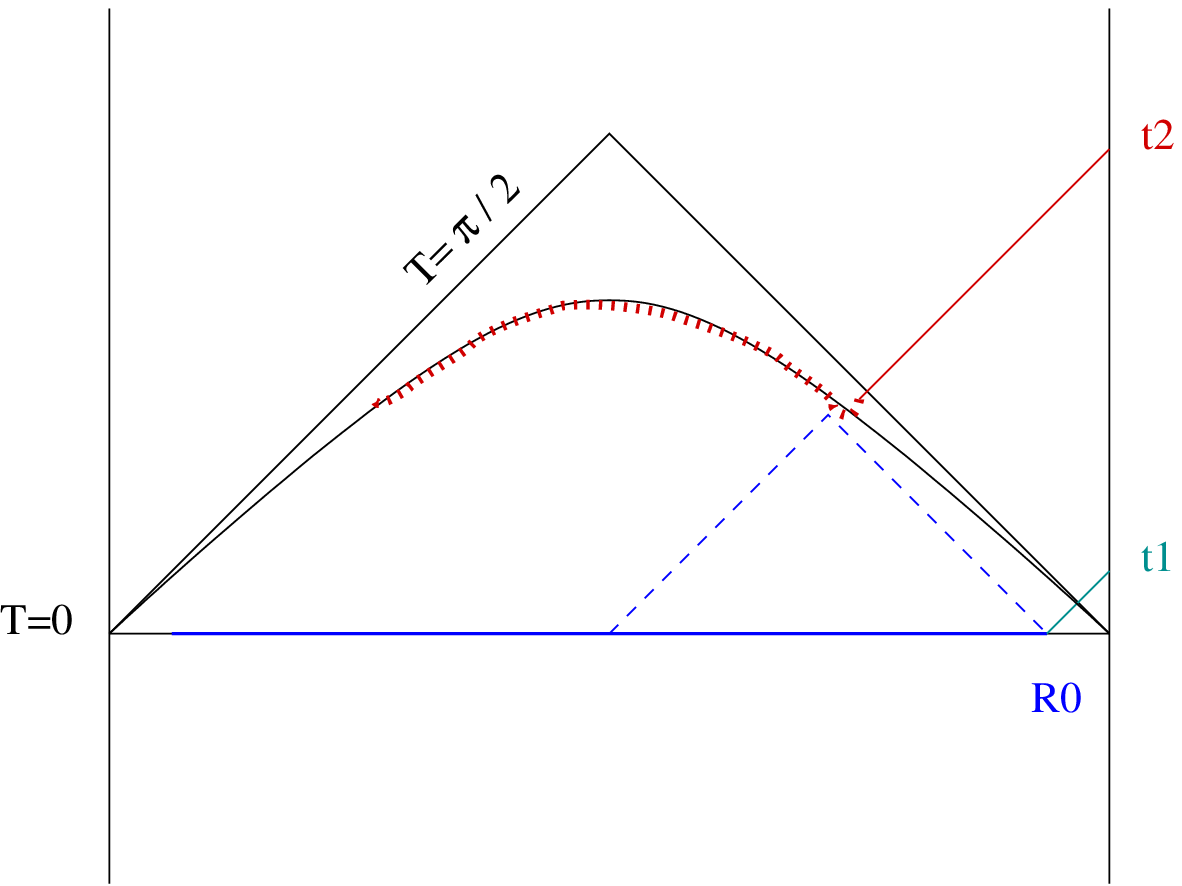}}

The ADM bulk energy is not conserved because of the slow fall off
of $\phi$. Only the CFT energy, which includes the boundary term
is conserved. Consequently, outside the domain of dependence of
the initial data the bulk energy could in principle get as large
as the CFT energy. One may think of this in terms of the extra
energy stored ``at the boundary" coming into the bulk. \pdns\
illustrates the following argument that the bulk energy can in
principle change on timescales shorter than that necessary for the
``formation'' of the black hole.

As reviewed in Appendix B,  the singularity forms on a (part of)
constant FRW time $T$ slice.  However, the singularity is
guaranteed to form only in the domain of dependence of the inner
region,  which requires $R_0$ to be quite large.  This in turn
implies that the propagation of the ``kink" in the initial field
profile \HHMphi\ (\ie\ the domain of influence of the inner
region) reaches the boundary at fairly early times (denoted $t_1$
in the diagram). So at $t_1$, we would expect the bulk energy to
start changing. Indeed, numerical evolution \kraus\ of the data
\HHMphi\  shows that the bulk energy starts to increase on short
timescales, rising quickly to the CFT energy value.

We cannot  easily determine the behavior of the singularity
outside the domain of dependence of the inner ($r < R_0$) region;
the singularity could in  principle turn around to become almost
null (or even timelike). But that means that to have a black hole,
with everywhere spacelike singularity, we only need to ``collapse"
it much later; in pure AdS this would be at time $t_2 \gg t_1$ as
sketched.  Conversely, in order to be certain that a naked
singularity will be produced, one would have to argue that the ADM
energy doesn't change (or at least become too large) by time
$t_2$. Actually, the situation is even worse, since if the black
hole did form, and on the causal diagram the singularity reached
the boundary at $t_2$, then from the boundary point of view $t_2$
would really be infinite.

To say this another way,  given that there is enough CFT energy to
make the required black hole, to motivate naked singularity we
would have to show that the black hole forms much ``later'' than
the singularity. Qualitatively, we expect the characteristic time
for singularity formation (from weak initial fields) should be of
order the AdS time. The time for black hole formation should be of
a similar order.  So there is no parametric separation in time
scales that would guarantee singularity formation well before
black hole formation.  For very large energy initial conditions,
like \HHMphi\ at large $\phi_0$, we expect both time scales to
shorten in a way which is unfavorable for the formation of naked
singularities. In particular, for strong initial fields and large
inner region the CFT time to see a naked singularity would be
$\sim { 1 \over R_0}$. On the other hand, if a large black hole
did form, we would expect it to settle down (and the CFT to
thermalize) on timescales given by the quasinormal mode
frequencies, which scale with the black hole temperature \qnm, and
therefore are given by\foot{ As we show later, the total energy in
the configuration \HHMphi\ is approximately $R_0^4 \, \po^2$.  If
a black hole forms, as is kinematically favorable, its mass is
likely to be the full mass of the configuration, so that its
horizon would be roughly $r_+ \sim R_0 \, \po^{1/2}$. Since this
is a very large black hole, the corresponding black hole
temperature would scale also as $T \sim R_0 \, \po^{1/2}$.}
 $\sim { 1 \over R_0 \, \po^{1/2}} \ll { 1 \over R_0}$.
Although the quasinormal modes represent the late-time behavior of
the black hole, given the parametric separation in the time
scales, this argument suggests that even if the singularity
becomes locally timelike, it would still be cloaked by the
horizon. As we shall see, there do exist large $\phi_0$
configurations of bounded energy.  For these configurations we
expect all time scales to be $\sim 1/R_0$.

 We have seen that the bulk energy
(which is initially too small to form a black hole) does not
necessarily remain small long enough to forbid black hole
formation.  The kinematic argument for naked singularities would
still apply, though, if the conserved CFT energy were too small to
make a sufficiently large black hole\foot{ On general grounds we
expect the CFT to thermalize eventually, resulting in a large
black hole in the bulk. So we expect any naked singularity to
eventually retreat behind the black hole horizon.}. But, as we
describe below, this energy appears sufficient to create the
required black hole. Here we review this result in a regulated
context\foot{The following argument was motivated by a discussion
with Juan Maldacena.}.

To understand this issue better, it is helpful to envision a
sequence of initial data $\phi_{\epsilon}(r)$ that approach the
configuration \HHMphi\ as $\epsilon \rightarrow 0$ and differ
significantly from it only at larger and larger $r$. Explicitly,
take
\eqn\phiep{ \phi_{\epsilon}(r)  = \phi_0 \equiv { A \over R_0^{2 +
\ep}} \qquad r \le R_0 \ , \qquad \;\;\;\; \phi_\epsilon(r)= {A
\over r^{2 +\ep}} \qquad r > R_0 \ . }
For finite $\epsilon$, the fall off at large $r$ is fast enough
that the bulk ADM energy coincides with the CFT energy and is
conserved.  We can imagine embedding these configurations in the
boundary CFT  with an effective UV CFT cutoff.  Field theory
intuition tells us that any physically meaningful phenomenon in
the boundary CFT should be understandable as the limit of the
behavior of a sequence of UV cut-off CFTs.  So if a naked
singularity is formed it should form for finite but small
$\epsilon$ and continue smoothly to $\epsilon = 0$.

The ADM (equal to CFT) energy of \phiep\ can be evaluated by using
the constraint equations \constraints. The detailed calculation is
outlined in Appendix A. As explained there, the $\ep \to 0$ limit
reveals a manifestly positive answer for the ADM mass.  We quote
here the dominant piece for small $\phi_0$;
\eqn\massepexp{\eqalign{ \Mc &= \pi^2 \, \phi_0^2 \, R_0^4 +
\cdots \ . }}

Since this energy is conserved and represents the largest possible
value for the bulk energy (because the boundary term is explicitly
positive), we can now ask whether this energy suffices to form a
black hole.  In AdS$_5$ a Schwarzschild black hole with horizon
size $\rp$ (in AdS units) has mass
\eqn\mbh{M_{BH} = 3 \pi^2 \, \rp^2 \, (\rp^2 +1 ) \ .}

We need to know the minimal Schwarzschild radius $\rs$ that will
enclose the singularity.  (By the area theorem, the size of the
event horizon of a black hole cannot shrink, so the final black
hole can only get bigger: $\rp \ge \rs$.) A simple ray tracing
argument as in \hhm, reveals\foot{ An argument of this type is
presented in Appendix B.} that one needs a Schwarzschild radius
\eqn\rsa{\rs > R_0 \, \phi_0^{2/3} \ , \qquad {\rm for} \;\;\; \po
\ll 1 \ .}
Evaluating at the lower bound,
\eqn\masshhm{ \Mbh > 3 \, \pi^2 \,  \phi_0^{8/3} R_0^4 \ , }
where we have restricted ourselves to the most favorable region, $
\rs \gg 1$.

Denoting
\eqn\mudef{ \mu \equiv {\Mbh \over \Mc} \ ,}
we see that in order to argue for the formation of a naked
singularity, we must show that $\mu > 1$. This indicates that the
configuration does not  have enough energy to collapse into a
black hole which would be sufficiently large to cloak the
singularity. Comparing \massepexp\ with \masshhm, we obtain
\eqn\massratio{ \mu >  3 \, \phi_0^{2/3}}
In the region where \rsa\ is applicable the lower bound in
\massratio\ gives $\mu \ll 1$ and the black holes are
energetically allowed to form.

The lower bound \rsa\ is clearly crucial in establishing possible
situations where naked singularities can form. We have explored
several ways to improve \rsa. The best result we have found makes
use of Raychaudhuri's equation for geodesic congruences.

Consider an outgoing congruence of null geodesics emanating from
the initial data surface. If the size of this congruence (the
proper size of the spheres at the given $r$ and $t$) starts to
decrease along the null geodesics, then Raychaudhuri's equation
shows\foot{This relies on the fact that the scalar field stress
tensor satisfies the null energy condition, despite the potential
\HHMpot\ being negative definite.}
 that the geodesics must terminate at a finite affine
parameter, ending in a singularity\foot{In general, the geodesics
could merely caustic; but for the spherical congruence we are
considering this would require the sphere to shrink to zero size,
which in the present context would still imply a singularity.}.
Now study outgoing null geodesics starting from larger and larger
$r$ at $t=0$.  If the proper size reaches a maximum at a finite
affine parameter (indicating the formation of a trapped surface),
then these geodesics must terminate at a singularity. Geodesics
that do not enter the trapped region are likely to continue to
expand and reach the boundary. Hence for some critical starting
radius there will be a geodesic congruence which skirts the
boundary of the trapped region within the homogeneous domain of
dependence where it can be easily calculated. The largest proper
size of this congruence within the homogeneous domain of
dependence will give a lower bound for the size of the
singularity. The details of this estimate are discussed in
Appendix B.5. Numerical solution of the homogeneous FRW equations
for $\phi_0 \ll 1$ gives the result
\eqn\rsb{ \rs > 0.58 \,  R_0 \, \po^{1/2} \ , \qquad {\rm for}
\;\;\; \phi_0 \ll 1 \ . }

For small $\phi_0$, \rsb\ is parametrically larger than \rsa, and
hence this will make $\mu$ much larger than before. Specifically,
computing $\mu$ with this estimate yields
\eqn\massratiob{ \mu >  0.34}
This is independent of $\phi_0$ for $\phi_0 \ll 1$ and is rather
close to unity!   The black hole barely has enough energy to form.
If the true size of the singular region $\rs$ were required by the
dynamics to be merely $ \sim 30\%$ larger than the bound \rsb ,
then naked singularities would be required to form.  Numerical
simulations of \frolov\ suggest that, at least in the regimes
considered, this does not happen: the size of the singularity
roughly matches our lower bound estimates.

At large $\phi_0$, the estimate for $\rs$ saturates at (\cf\
Appendix B.5)
\eqn\rsc{\rs> \ss \, R_0 = 0.577 R_0 \ , \qquad {\rm for} \;\;\;
\phi_0 \gg 1~ .}
For the initial configuration \HHMphi, the large $\phi_0$ energy
is (see Appendix A.1, Eq.\masstwolarge)
\eqn\mtl{M_c= 10.2 \; \po^2 \, R_0^4  \ ,  \qquad{\rm for} \;\;\;
\po \gg 1 \ .}
Computing the mass ratio, we find
\eqn\mularge{\mu > {0.32 \over \po^2} \ ,}
which is much smaller than one.

The configuration \HHMphi\ is close to the lowest energy
configuration subject to boundary conditions for $\phi_0 \ll 1$.
However, this is no longer the case for large $\phi_0$ (despite
${1 \over r^2}$ being the slowest normalizable
 asymptotic fall-off).
Since we would like to find a regime wherein we could argue that
the initial configuration is too light to form a black hole
cloaking the singularity, it is clearly desirable to start with
the lowest possible energy configuration for given $R_0$ and
$\po$. Hence we now consider the situation when $\po \gg 1$ more
carefully. In this case we see from \rsc\ that the size of the
black hole that needs to form is independent of $\po$. In choosing
\HHMphi\ we are paying an enormous price in terms of energy for
large $\po$, see \mtl. However, we can construct configurations
whose energies are independent of $\po$ very easily. Instead of
having the scalar field drop to zero very slowly as in \HHMphi, we
pick a rapid fall-off, say
\eqn\nphi{ \phi(r)  = \phi_0 \equiv { A \over R_0^n}  \qquad r \le
R_0 \ , \qquad \;\;\;\; \phi(r)= {A \over r^n} \qquad r > R_0 \ .
}
The calculation of the mass for this profile is outlined in
Appendix A. We find that taking $n \to \infty$, with $\po$ fixed,
yields
\eqn\stepmass{ \Mc = 3 \, \pi^2 \, R_0^2 \, (1+R_0^2) \ , }
which is the mass of a AdS-Schwarzschild black hole with size
$R_0$ (this statement is true for all values of $\po$) \foot{The
strict $n\to \infty$ limit results in the field profile being a
step function. This case can be visualized as a thin domain wall
junction between an AdS-Schwarzschild black hole of size $R_0$ and
a FRW cosmology with scale factor set by $\po$; a generalization
of the ``Wheeler bag of gold'' to asymptotically AdS spacetimes.}.
In the large field limit making use of \rsc, we see that
\eqn\phiindepmu{ \mu ={ \rs^2 ( 1 + \rs^2) \over R_0^2 (1+ R_0^2)}
\sim (0.58)^4 = 0.111 }

This step function configuration may not be the minimum energy
configuration for large $\phi_0$. However, it illustrates the
important fact that even for strong fields black holes are not
parametrically lighter, unlike the case with \HHMphi. It might
indeed be possible that a case can be made for naked singularity
formation by using the true minimum energy configuration along
with a tighter lower bound than \rsb\ and \rsc\ for the size of
the requisite black hole.

Hence, we see that initial $\phi$ profiles  containing a flat
region to allow study of singularity formation, favor formation of
black holes big enough to cloak singular regions (of the lower
bound size we established) in all regimes we could analyze.
However, this does not necessarily imply that the exact dynamical
evolution of the initial data will not create a naked singularity.
One way this can be studied is  by a careful numerical evolution
of the data. Significant progress in such numerical evolution has
been made in \kraus, though the regime of $\rs$ they used is not
the most optimal for forming naked singularities.

Given that the 5-dimensional setup discussed above did not yield
kinematic proof for formation of naked singularity, one might hope
that we could do better in a different number of dimensions.
Unfortunately this does not seem to be the case (see Appendix B).

\newsec{AdS/CFT with a cut-off}

In the previous section we saw that the problem with using the
bulk ADM energy for \HHMphi\ (which was negative) was its
non-conservation. The conserved energy did not prevent the
formation of black holes that cloak the singularity. A different
possibility raised by HHM \hhm (v2), is to consider a cut-off AdS
geometry where one imposes boundary conditions for the fields at
the cut-off surface $r=R_1$ for all time. Then the bulk ADM energy
is conserved, and one can use it to study the possibility of
forming naked singularities.

\subsec{Weak fields in the cut-off geometry}
As in \hhm, we consider the scalar-gravity system \HHMaction, in
AdS$_5$ with\foot{To ensure that we have a sufficiently large
region of AdS geometry, we choose $R_1 \gg R_0 \gg 1$.} $ r \le
R_1$ and fix $\phi(R_1)= \phi_1$ for all time.  We will attempt to
find solutions where $\phi(r) \ll 1$ so a linearized analysis will
be valid.   Since $\phi(r)$ has Dirichlet boundary conditions at
finite $r$, both the normalizable mode of the linearized field
equation, $\phi(r) \sim 1/r^2$, and the non-normalizable mode,
$\phi(r) \sim \log(r)/r^2$ contribute. The non-normalizable mode
approaches the boundary more slowly and, roughly speaking, has
less energy than the normalizable mode.   To begin we choose
$\phi_0$ and $\phi_1$ to pick out just the non-normalizable mode%
\foot{Note that this or any configuration does not smoothly allow
the cut off to be taken away, unlike the situation discussed in
the previous section.} {\it cf.}, \hhm:
\eqn\cutconf{\eqalign{ \phi(r) &= \phi_0 \ , \qquad r \le R_0 \cr
\phi(r) & = {\phi_1 \, R_1^2 \over \ln R_1} \;\; {\ln r \over r^2}
\ , \qquad R_0 <r \le R_1 \ . }}

The ADM mass $\Mc$ of the  configuration \cutconf\ can be computed
using the constraint equation \mdiffeq. For $\phi_0 \ll 1$  we
find
\eqn\masscut{
 \Mc = 2 \pi^2 \, \po^2 \, R_0^4 \,
\( - \( {\ln R_1 \over \ln R_0} \)^2
   + {\ln R_1 \over  2 \, (\ln R_0)^2 }
   + {1 \over 2} \) }

We should compare this with the energy required to form a black
hole of size $\rs$. Due to the new boundary conditions, the black
hole in question is not the usual Schwarzschild-AdS black hole,
(since the Schwarzschild-AdS black hole would not admit static
scalar field outside the horizon), but rather an AdS scalar hair
black hole \torii. For weak fields outside the black hole we can
use a linearized approximation and ignore backreaction. The
boundary conditions on the scalar field are imposed at $R_1$,
$\phi(R_1) = \phi_1$ and at the horizon $r=\rs$. From the field
equations \scbheqns\ and the fact that $f(\rs)=0$ we derive the
linearized horizon boundary condition $ \phi(\rs) + \rs \,
\phi'(\rs) = 0 $. A linear combination of the two modes that
satisfies the boundary conditions is
\eqn\bhphia{ \phi_{bh}(r) = \a_{bh} \; {1 \over r^2 } \, \ln\({r
\, e \over \rs}\)\ , }
where $\a_{bh}$ is determined by the boundary condition at $R_1$.
We find (for convenience taking $R_1  \gg R_0 \gg \rs \gg 1$)
\eqn\bhphib{ \a_{bh}  = {\phi_0 \, R_0^2  \over \ln R_0} \, \(1 -
{\ln (\rs/e)  \over \ln R_1 } \)^{-1} \ . }
Note that the coefficient \bhphib\ of the $\ln(r) /r^2$ fall off
is slightly larger than in \cutconf.  This means that the black
hole with the same $\phi_1$  has a little more scalar hair than
the initial configuration.  The black hole field profile is not
required to go through $\phi_0$ at $R_0$ and so can become larger
to minimize its energy.   We see this by computing the black hole
energy. As explained in Appendix A.3, we find
\eqn\msbh{\eqalign{ M_{SBH} & = 3 \, \pi^2 \, \rs^2 (1+\rs^2) \cr
& \qquad + 2\,  \pi^2 \, \po^2 \,R_0^4 \, \(1 - {\ln (\rs/e)
\over \ln R_1 } \)^{-2} \,
\[- \( {\ln (e \,R_1/\rs) \over \ln R_0 } \)^2 + { \ln R_1\over 2 \,
(\ln R_0)^2} \] . }}
This reduces in the large $R_1$ limit to
\eqn\msbhaa{
 M_{SBH}  = 3 \, \pi^2 \,  \rs^2 (1+\rs^2)  +
2 \, \pi^2 \, \po^2 \, R_0^4 \, \(  - \( {\ln R_1 \over \ln R_0}
\)^2
   + {\ln R_1 \over  2 \, (\ln R_0)^2} \) \ .
}
Using the previously derived estimate \rsb\ for the minimum
Schwarzschild radius necessary to cloak the singularity we find
\eqn\sbhmdiff{ M_{SBH} - \Mc = - 0.66 \, \pi^2 \po^2 \, R_0^4 \, }
So we see that black holes which can cloak the singularity cannot
be excluded from forming\foot{Our results differ from those in
\hhm (v2) because we employ the optimal scalar hair
configuration.}. To check the validity of the linearized
approximation we compute $\phi(\rs) = 1/\log R_0 \ll 1$.   So we
can have a large singular region and a controlled linearized
analysis at the same time.  This would not be the case with the
weaker lower bound \rsa\ .

 We have studied other initial $\phi$ profiles within
the linearized regime and have not found a case where naked
singularities were required to form. The well behaved nature of
weak field initial conditions makes us suspect that they are not
the best candidates for finding a violation of these lower bound
estimates.  So we now turn to the nonlinear regime where $\phi$
need not be small.

\subsec{Minimal energy configurations at strong field}

To find a minimum energy initial configuration in general we need
to minimize the mass function $m(r)$ with respect to the profile
$\phi(r)$. While it is not difficult to derive the Euler-Lagrange
equations (by taking a variation of \massfn), it helps to think of
the coupled gravity-scalar system  minimize the Einstein-Hilbert
action (see Appendix A.2). To specify the problem completely we
need boundary conditions; we will require that we have a
homogeneous central region of size $R_0$ and $\phi(R_0) = \po$.
This determines $m(R_0)$ for us; the mass is just the contribution
from the effective cosmological constant. Our final boundary
condition is the Dirichlet boundary condition at the cut-off
surface $\phi(R_1) = \phi_1$. So specification of $(R_0, \po, R_1,
\phi_1)$ along with the Einstein's equations for static,
spherically symmetric profiles is a complete specification of the
minimization problem. It should be no surprise that the equations
of motion for the initial configuration are identical to that of
the scalar hair black hole (which of course has different boundary
conditions).

We find the dynamics simplifies in the strong field regime,
$\phi_0 \gg 1$. In this limit the coupled equations have a
universal attractor solution  $\phi_{att}(r) = \sqrt{3} \log
({R_1/r}) + \phi_1$ that obeys the boundary condition at $R_1$.
This means that the minimum energy initial configuration and the
scalar hair configuration have the same functional form for $r >
R_0$.  In particular, the scalar hair outside the black hole
horizon $\rs$ has the form $\phi_{sbh}(r) = \phi_{att}(r)$, $ r
>\rs$. The initial profile is given by
\eqn\nlcutconf{ \phi(r) = \phi_0 = \phi_{sbh}(R_0) \ , \qquad r
\le R_0  \ , \qquad \& \qquad \phi(r) = \phi_{sbh}(r) \, \qquad
R_0 <r \le R_1 \ . }
We find in Appendix A.2 that the mass difference between this
optimal initial configuration and the scalar hair black hole of
size given by the bound $\rs > \ss R_0$ with $\ss = .577$
(equation \rsc), is given by\foot{This estimate assumes, for
simplicity, that $R_0, R_1$ are fixed and $\phi_0 \rightarrow
\infty$.}
\eqn\nlmassdiff{ M_{SBH} - \Mc =   {\pi^2 \over 3} \, \ss^2 \,
R_0^4 \, (4 \, \ss^3 - 1) \;\;
 \,  e^{{2 \over
\sqrt{3}} \ps } \, e^{-{1 \over 3} \,  \int_{R_0}^{R_1} \, x \,
\phi_{sbh}'(x)^2} \;\; < \;\; 0 \ .  }

Again naked singularities are not required to form. One can see
that the critical value of $\ss$ for guaranteeing naked
singularity formation is $\ss^* = {1\over 4^{1/3}} \approx 0.63$.
This implies that only a  10\% increase in our estimate for the
size of the black hole would be sufficient to rule out its
formation on kinematic grounds.

\subsec{Field theory perspective}

So far we have focused on the 5D cut-off supergravity system to
ask whether naked singularities appear to form generically. But in
the full 10D theory new issues emerge. This cut-off theory turns
out to be unstable\foot{We thank Ofer Aharony and Matt Strassler
for enlightening us on this point.} (see for example, \bce).
First, note that \cutconf\ (and others like \nlcutconf) correspond
to deforming the boundary theory by a relevant operator of mass
dimension $\Delta = 2$; $\CO_2 = {\rm Tr}\, ( X^i X^j) - {1\over
6} \delta^{ij} \, {\rm Tr} \, (X^2)$, where $X^i$ are the adjoint
scalars in the $\CN =4$ Super Yang-Mills multiplet, transforming
in the vector representation of the R-symmetry group $SO(6)$. The
operator $\CO_2$ is in the traceless symmetric two-tensor
representation of $SO(6)$. Because of the symmetrization and trace
removal, this operator has negative eigenvalues. These negative
eigenvalues induce tachyonic mass terms for the adjoint scalars.
Hence $\CN =4$ SYM on  $\R^4$ deformed by $\CO_2$ is unstable.

We are interested in the field theory on $\S^3 \times \R$, since
we wish to work with global coordinates. There is a crucial
distinction here arising from the curvature $\CR$ of the boundary.
The $\CN =4$ Hamiltonian on $\S^3 \times \R $ has positive mass
terms for the adjoint scalars from the conformal couplings to the
background curvature. In particular, the Hamiltonian for the
deformed theory is schematically
\eqn\nfourham{ \CH  = \CH \(\R^4 \) + \CR \, {\rm Tr} \,(X^2) +
\hat{\a} \, \CO_2 }
We see that the theory defined by $\CH$ is stable as long as $\a =
\hat{\a}/\CR < 1$. We can read off $\a$ from the bulk initial
$\phi$ configuration at large $r$.  It is just the coefficient of
the $\log r/r^2$ asymptotic fall-off.    For optimal weak field
scalar profile \cutconf, we find $\a = \po \, R_0^2 /\ln R_0$. For
any fixed $\phi_0 \ll 1$ as the size of the flat region $R_0$
grows $\a$ becomes large and the theory becomes unstable. There
are very weak field configurations with finite singular regions
that do remain stable, though. Using the lower bound \rsb\ we can
write $\a \sim \rs^2/\ln R_0$ which can be made small for fixed
large $\rs$. The generic strong field configuration is also
unstable, although the profile can be fine tuned to make $\a$
vanish.  So overall, it seems likely to us that the best
possibilities for finding naked singularities will have unstable
CFT duals.

\subsec{Supergravity manifestation of CFT instability}

Our argument for the instability of the gauge theory defined in
\nfourham, was based on our expectation from perturbative gauge
theory. Understanding what this means for the dual supergravity
theory is quite instructive\foot{  We have already seen that $\CN
= 4$ SYM on manifolds with non-trivial curvature has adjoint
scalars conformally coupled to the background curvature. Formally,
the Hamiltonian \nfourham\ is analogous to the case where the
manifold has negative curvature. In the latter case that analysis
of \sw\ tells us that there is a non-zero probability for
nucleating BPS D3-branes which are pushed off to the boundary of
AdS.}.

We expect that the theory deformed by \nfourham\ has a dual
supergravity geometry, where BPS D3-branes have a potential that
drives them to the boundary of the AdS geometry. For $\CN =4$ SYM
deformed by $\CO_2 $ on $\R^4$, this was studied in \bce, where
bulk equations were solved to find the supergravity dual.
Furthermore, the probe D3-brane potential was evaluated and shown
to be unbounded from below.  We now extend this analysis to global
coordinates.

To compute the probe potential, we need to know the full
ten-dimensional  geometry that is dual to the deformed field
theory. Thus far we have been  dealing with five-dimensional
gauged supergravity with a single  scalar field. Happily for us,
there exists a unique lift of  five dimensional solutions to
ten-dimensions, via the magic  of consistent truncation \clpst. In
practice we need to consider a static, spherically symmetric
ansatz for the five-dimensional metric and solve the  coupled
scalar-gravity system of \HHMaction. However, to get  insight into
the probe potential, it suffices to work  in a linearized
approximation about the AdS background.  This lets us obtain the
potential felt by a D3-brane, whose  world-volume is $\S^3 \times
\R$, as a function of the radial coordinate $r$ and the directions
on the $\S^5$. Parameterizing the $\S^5$ as in \sf\ we find (see
Appendix C for details):
\eqn\effpot{ V_{\rm probe} = {1\over 2} r^2 \, \( 1 - {\alpha
\over 2 \sqrt{3} } \, \( 3 \, \sin^2 \xi  -1 \) \) \ , \qquad {\rm
for} \;\; r \gg 1  \ .  }
In deriving \effpot, we assume that the supersymmetric D3-brane
saturates the BPS bound.

It is not surprising that this potential is identical to the
tree-level potential in the SYM action because large radius
corresponds to working near the UV fixed point of the boundary
field theory. As advertised, there is a critical value $\a_c = 2
\sqrt{3}$ above which the probe potential is unbounded below. We
have computed the asymptotic form of the potential; it is possible
that the true potential has a local minimum in the interior of the
geometry. This however will not help stabilize the theory in the
high energy regime, as the branes can be excited out of the well
to infinity. The behavior of probe branes in the dual supergravity
justifies our fears about the CFT being ill-defined in the regime
$\a > \a_c$.

Apart from this instability, there is general concern with the
above analysis. In the context of AdS/CFT  radial evolution
corresponds to  RG evolution in the boundary theory. In
particular, moving toward the center of AdS (\ie, small $r$)
corresponds to the IR of the dual field theory. Perturbations of
any field theory by a relevant operator grow in the IR, possibly
causing drastic modifications, which are reflected in the
supergravity dual. In many examples one manifestation of these
effects is a static naked singularity  at finite $r$ in the 5D
truncated gravity theory ({\it cf.}, \bce ). But in 10D the
effects often take a different form, often involving brane
dynamics (see for example \fpw, \PolchinskiUF). The breakdown of
the linearized approximation in the 5D supergravity indicates that
the effect of $\CO_2$ are getting large.  So there may well be new
dynamical channels available for the behavior of the 10D theory.
In order to fully understand its behavior these effects must be
incorporated.

\newsec{A stable example in  3 dimensions}
It is important to know if there are any examples where the
underlying theory is stable and naked singularities are produced.
It appears there may be such an example in $d=3$.

 For scalar fields that saturate the BF bound, the story is very
similar in all dimensions $d \ge 4$. In  supergravity backgrounds
of the form AdS$_d \times \S^q$, there are  scalar fields
saturating the BF bound $m^2_{BF} = -{(d-1)^2 \over 4}$, for odd
$q>3$ \dfghm. This scalar field is a particular linear combination
of the graviton mode and the RR-field, polarized in the $\S^q$,
with its wavefunction being a spherical harmonic  of angular
momentum $l= (q-1)/2$ on $\S^q$. Since the spherical harmonics are
non-positive definite, the operator dual to the field has negative
eigenvalues. Similar statements can be made if we replace the
$\S^q$ by some other compact Einstein manifold. In all these
situations the dual CFT will suffer from an instability.

The situation in AdS$_3$ is however very different. For
concreteness we will concentrate on the case of AdS$_3 \times \S^3
\times X$, where $X$ may be $T^4$ or $K3$. In this case the BF
scalars do not arise from modes of the graviton on the $\S^3$.
They come from KK-reduction of anti-self dual tensor fields in
six-dimensional supergravity\foot{ Type IIB theory on $X$ is
described by a six dimensional low energy effective lagrangian,
whose field content is the super-graviton multiplet and $n$ tensor
multiplets, with $n=5$ for $T^4$ compactification and $n=21$ for
$K3$. The anti self-dual tensor fields are part of the tensor
multiplet.} \ms, \dkss. These scalars are expected to be dual to
fermion bilinears with $\Delta = 1$ in the dual CFT \ms.

Since the dual operators are fermion bilinears, we might hope that
the deformed boundary theory is stable. But, it is plausible that
the nature of the anti self-dual tensor field's wavefunction on
the $\S^3$ (it is a $l=1$ spherical harmonic) may result in an
instability, which in this case would be manifested by nucleating
BPS D-strings. Nonetheless, there is a way to stabilize the theory
on the boundary. We can consider the boundary CFT  at a point in
moduli space ({\it cf.}
 \dijk, \sw\ for discussions of the moduli space) where the
marginally bound BPS objects are lifted from the spectrum.

Recall that the geometry AdS$_3 \times \S^3 \times X$ is the
near-horizon geometry of the D1-D5 system, with the D5-branes
wrapping $X$. If we turn on NS-NS fluxes through $X$, then we will
resolve the small instanton singularity and make the D1-D5 system
truly bound, as opposed to marginally bound as in the absence of
the fluxes. Then a probe D-string feels a force \dmwy, which
results in a positive definite potential. The deformation caused
by the operator adds a subleading contribution which could be
negative, but the net effect would be such as to stabilize the
system.

Schematically, the potential will now look like
\eqn\donepot{ V_{\rm probe} = B \, r^2 - \a \, \log(r) \ , \qquad
{\rm for} \;\; r \gg 1 \ . }
The leading term $B \, r^2$ is a result of turning on fluxes
through the internal space. In case of the probe D3-brane
calculation outlined in Appendix C, this term maps to the $r^4$
contribution that canceled between the DBI and the WZ
contributions to the potential. The subleading term is
proportional to $\a$ (upto order 1 coefficients) and is a direct
consequence of the deformation of the CFT. Note that in this case
the boundary CFT lives on $\R \times \S^1 $, the boundary of
AdS$_3$ and hence we do not get a mitigating contribution from the
curvature coupling. We note in passing that a very similar ideas
were explored in \mm, in the context of multi-boundary
Euclidean-AdS geometries.

We are now in a position to propose a scenario which does not
suffer from instabilities.  The set-up we would like to propose
has 3-dimensional gravity coupled to a single scalar field $\phi$,
with some scalar potential $V_3(\phi)$.
\eqn\threedlag{ S_3 = \int \, d^3 x \, \sqrt{g} \, \( {1\over 2}
\, R - {1 \over 2} \, (\nabla \phi)^2 - V_3(\phi) \) }
The exact scalar potential $V_3(\phi)$ can in principle be
extracted from \nics. For our purposes a quadratic approximation
will suffice: $V_3(\phi) = -1 - {1\over 2} \, \phi^2 $. Ideally we
should check that the scalar $\phi$ does not source other three
dimensional fields to ensure consistency of the truncated
lagrangian \threedlag. We will assume that this is true.

For a scalar of mass $m^2 = m_{BF}^2 = -1$,  the Klein-Gordon
equation on AdS$_3$ background  tells us that $\phi(r) \sim \,
C_1/r + C_2 \, \log(r)/r$ at large $r$.  The arguments presented
in Section 3 will continue to apply in  this case as mentioned
before. Configurations analogous to \cutconf\ and \nlcutconf\ will
as in five dimensions continue to have enough energy for the
formation of black holes. As in those cases it may be possible to
rule out black hole formation by getting a more accurate lower
bound on its size.

This configuration is expected to be dual to  a well behaved
two-dimensional QFT.  From the form of the probe potential
\donepot, it is possible that there is a global minimum at some
finite $r$. This can be avoided by tuning the fluxes so that the
D-strings are stabilized at some small radial distance,  so that
there is definitely a regime in the domain of dependence  of the
region $r \le R_0$ on the initial data slice, where the formation
of a naked singularity is a possibility.

The value of the scalar field is large at $R_0$ here as well, so
the effect of the relevant operator is large.   So the boundary
CFT vacuum structure and the 10D bulk theory is dramatically
altered at small $R$ (in the IR).   These dramatic alterations may
alter the prospects for naked singularity formation in the full
10D theory.   One concrete effect we can point to is that the
relevant operator breaks supersymmetry, and so will induce a
potential for the various moduli.  This might make the values
chosen to stabilize the branes unstable.

\newsec{Discussion}

The history of cosmic censorship conjecture has been characterized
by many plausible counter-examples, most of which on close
examination have proved to be incorrect. The fascinating aspect of
the HHM proposal is that it is based on an extremely simple
kinematic argument, which essentially says that in the
configurations of interest there is not enough energy to cloak the
putative singularity by a black hole horizon.

The proposal of HHM for cosmic censorship violation in
asymptotically AdS spacetimes with general scalar potential is
attractive, but there remains a gap in the arguments  as we
illustrate numerically in Appendix B.6 and has been independently
discussed in \hhmrev . The HHM proposal in the AdS/CFT context is
particularly interesting, though, because of the potential for
learning about stringy singularity resolution.

Within 5D supergravity, HHM  discussed  two distinct scenarios to
probe the formation of naked singularities.  The first one, in
infinite volume AdS, suffers from the problem of the bulk energy
not being conserved, and the conserved energy being sufficient for
the formation of a black hole cloaking the singularity. This does
not necessarily mean that a black hole will form.  It is possible
that the lower bound on the size of the singular region used in
these estimates is significantly exceeded in the exact dynamics.
Or the singularity could occur before the black hole has a chance
to form.  Numerical evolution of the field equations is one
promising way to address these possibilities.   The numerical
study carried out recently in \kraus\ indicates that evolution of
the HHM data results in  the formation of a trapped surface,
pointing toward black hole horizon formation. This is suggestive,
but because trapped surfaces are ubiquitous in the central region
it will be important to follow the evolution further and see an
external black hole geometry form. Numerical studies are also
being carried out by \frolov\ (using coordinates that are regular
at the horizon) which suggest black hole formation; in fact, the
final size of the singularity is approximately equal to our
analytic lower bound estimate.

In the alternate scenario of cut-off AdS/CFT we have seen again
that black hole formation is kinematically possible and we cannot
argue that naked singularities must form.   It is possible that
they do form, though, for reasons mentioned above.

However, for the most likely initial data for singularity
formation, the full 10D theory with cut-off suffers from an
instability: D3 brane shells expand outward to the AdS boundary.
This is clear from the boundary gauge theory point of view and
from a bulk D-brane probe calculation.

Let us assume for a moment that a naked singularity does form in
the AdS/CFT context. Assuming the 5D supergravity is a consistent
truncation of 10D supergravity, we can lift the 5D naked
singularity solution to an exact naked singularity solution of 10D
supergravity. The central question would now become, what resolves
this singularity?  One possibility is that stringy effects do so.
This is presumably the case for the non-generic Choptuik
\choptuik\ type naked singularities.  These are, roughly speaking,
the threshold formation of an arbitrarily small Schwarzschild
black hole.  This should be resolved in string theory by the
correspondence between very small black holes and highly excited
strings \SusskindWS, \HorowitzNW.  It is clear that any such
resolution of the singularities discussed by HHM will involve the
dynamics of D3-branes in 10D.  The work of \fpw, \bce\ make clear
that the static naked singularities in these 5D supergravities are
dimensionally reduced images of D3-brane distributions in 10D.
More generally we have come to expect naked singularities in lower
dimensional theories to have explanations in terms of brane
dynamics in the full 10D theory \JohnsonQT, \PolchinskiUF.

There are other issues to consider in the cut-off HHM context. The
instability of this theory is also due to brane dynamics.  It
should be visible in the classical 10D supergravity equations
because black 3-branes with finite horizon size can be described
there\foot{There is a subtlety here. The black 3-brane of finite
horizon size is not BPS and so the stabilizing $r^4$ term in the
probe brane potential is present. But in the strong field regime
and horizon sizes of AdS scale the stable brane position will be
at $r \gg R_0$}. It is possible that with generic 10D initial data
such branes will be created and then move to the boundary\foot{The
behavior near the singularity will be modified, but not made
nonsingular, by the lift to 10D supergravity.   A transition to
chaotic BKL type behavior is expected \DamourWM.}. This will
decrease the energy and RR charge, resulting in a space of higher
and higher negative curvature, presumably ending in a completely
singular geometry. From the boundary CFT point of view, as
D-branes move to the boundary, the effective $N$ of the $SU(N)$
theory decreases, resulting in a smaller AdS radius. The endpoint
is $N \sim 1$. In any event, the true long time behavior of the
system will be dominated by this instability.

The other possibility is that stringy and/or quantum effects near
the singularity are required to create D-branes that then move to
the boundary.  The production would presumably continue until all
D-branes move to the boundary and the whole space becomes
singular.

Finally, we saw that there may well be a stable AdS$_3$/CFT$_2$
example. In this case we have argued that the rich moduli space of
the dual CFT may be put to use in stabilizing the theory. Here we
will have to numerically investigate the scenario to check whether
we have a good kinematic argument for naked singularity formation.
But here (as in the 5D case) the relevant perturbation will have a
large IR effect which could dramatically affect the discussion of
the fate of the singularity.

An interesting question which we have not addressed in this paper
concerns the signature of naked singularities that might be seen
in the dual gauge theory. A priori it seems clear that the
signatures ought to be more marked and easier to access than those
of the black hole singularities as studied in \fhks. The formative
stages of a naked singularity will be characterized by increase in
curvatures and local correlators on the boundary should be
sensitive to these effects. One might speculate that the formation
of the singularity would be characterized by some spiky behaviour
in the boundary correlators which would be smoothed out upon
inclusion of quantum effects.

A related question concerns the signals of an ``almost" naked
singularity. We have found a number of examples where if the
singularity is cloaked at all, the maximum proper time between the
horizon and the singularity will be much less than one AdS time.
  This almost naked singularity should have an
impressive signature to outside the horizon observers ---
presumably large curvature oscillations that eventually settle
down to the black hole geometry.   These oscillations should
contain large amounts of information about the singularity.

A probe of the kind discussed in \fhks\ may well be useful here,
at least for analytic initial data.  The signal of the behind the
horizon singularity discussed there was a pole on the second sheet
of a correlation function, a bit like a broad resonance.  Perhaps
as the black hole singularity comes closer and closer to the
horizon the resonance becomes narrower and narrower, possibly even
emerging from behind the horizon onto the first sheet.

\vskip 1cm \centerline{\bf Acknowledgments}

We would like to thank Ofer Aharony, Andrei Frolov, Michael
Gutperle, Thomas Hertog, Gary Horowitz, Petr Ho\v{r}ava, Shamit
Kachru, Per Kraus, Juan Maldacena, Simon Ross and Matt Strassler
for useful discussions. This work is supported in part by NSF
grants PHY-9870115 and PHY-0098840, DOE grant DE-AC03-76SF00098,
the Stanford Institute for Theoretical Physics and the Berkeley
Center for Theoretical Physics.


\appendix{A}{Masses of various configurations}
\subsec{Power law fall-off profiles in five dimensions}

Consider the lagrangian \HHMaction. We are interested in picking
some initial data slice, a spacelike hypersurface $\Sigma$ at
$t=0$, and evolve the data prescribed on this surface forward in
time. We choose time symmetric configurations which allow us to
set all time derivatives to zero. Einstein's equations projected
onto $\Sigma$ give us the constraint equations that the data is
required to satisfy. For \HHMaction\ the equations read:
\eqn\constraints{ ^{(4)}\CR  = g^{ij} \, \partial_i \phi \,
\partial_j \phi + 2 \, V(\phi)\ , }
where $g_{ij}$ is a Euclidean metric on the surface $\Sigma$ and
$^{(4)}\CR$ the associated Ricci scalar. For spherically symmetric
configurations we can parametrize the metric on $\Sigma$ as
\eqn\sigmamet{ ds^2_{\Sigma} = \(1 - {m(r)\over 3\, \pi^2 \, r^2 }
+ r^2 \)^{-1} dr^2 + r^2  \, d\Omega_3^2 \ . }
For this geometry the ADM mass is defined to be
\eqn\massadm{ \Mc = \lim_{r \to \infty} \, m(r) \ . }
 From the constraint equation \constraints, we find:
\eqn\mdiffeq{ m_{,r} + {1 \over 3} \, r \, \(\phi_{,r}\)^2 \, m =
2 \pi^2 \, r^3 \, \[V(\phi) + 6 + {1\over 2} (1+r^2)\,
\(\phi_{,r}\)^2
\] \ .
}
This can be integrated to give
\eqn\massfn{\eqalign{ m(r) &= 2 \pi^2 \, {1\over F(r)} \, \int_0^r
\,F(x) \; G(x) \,dx \cr F(x) &=  \exp \(\int_0^x \, dy \,{1\over
3} \, y \, (\phi_{,y})^2 \) \cr G(x) &= x^3 \,\[V(\phi) + 6 +
{1\over 2} (1 + x^2)\, (\phi_{,x})^2 \] }}

Before proceeding, let us write the potential \HHMpot\ in a more
useful form;
\eqn\sugrapot{\eqalign{ V(\phi) &= - 2 \( \, e^{{2 \over \sqrt{3}
} \, \phi} + 2
  e^{-{1\over \sqrt{3} } \, \phi} \)
= \sum_{m =0}^{\infty} \, a_m \,  \phi^m \cr {\rm where} \qquad
a_m & = - { 2 \over m!} \, {1 \over 3^{{m \over 2}}} \, \(2^m +2
(-1)^m \) }}
Consider a general field profile,
\eqn\genphidef{ \phi(r)  = \phi_0 \equiv { A \over R_0^n}  \qquad
r \le R_0 \ , \qquad \;\;\;\; \phi(r)= {A \over r^n} \qquad r >
R_0 \ . }
We then have
\eqn\Fdef{ F(x) = 1 \qquad x \le R_0 \ ,  \qquad \;\;\; F(x)=
\exp\( -{\zeta \over x^{2n}} + {\zeta  \over R_0^{2n}} \) \qquad x
> R_0 \ , }
with
\eqn\aldef{
 \zeta \equiv {n \over 6}\, A^2  \ ,
}
and
\eqn\Gdef{\eqalign{ G(x) & = x^3 ( V(\phi_0) + 6)       \qquad x
\le R_0 \ , \cr & = x^3\(  \sum_{m =2}^{\infty} \, a_m \, {A^m
\over x^{m \, n}} + {n^2 \, A^2 \over 2} \, (1+x^2) \,{1 \over
x^{2n+2}}\) \qquad x > R_0 \ . }}

The contribution to $G(x)$ from the $m=2$ term in the summation
and the last term in \Gdef\ have scale together as $x^{3 - 2 n }$,
with a coefficient $A^2 \(-2 + { n^2 \over 2} \) $. This vanishes
at  $n =2$, but closer examination reveals that the integral is
divergent and in fact behaves as $1/(n-2)$. To see this, it is
useful to introduce the incomplete Gamma function:
\eqn\gaminc{ \gamma(\beta, x) = \int_0^x \, dt \, e^{-t} \, t^{-1
+ \beta} }
This is strictly defined for ${\rm Re}(\beta) >0$. The $\beta \to
0$ limit is defined using confluent Hypergeometric function and
one has:
\eqn\gamlim{ \lim_{\beta \to 0} \, \beta \, \gamma(\beta,x) = 1 }

Evaluation of \massfn\ for \genphidef\ is straightforward :
\eqn\adm{\eqalign{ {1 \over 2 \pi^2} \,  \Mc &= - {R_0^4 \over 2}
\, \[ \exp\( 2 \sqrt{{2 \over n}} \, {\sqrt{\zeta} \over R_0^n} \)
+ 2 \, \exp\( -\sqrt{{2 \over n}} \, {\sqrt{\zeta} \over R_0^n} \)
 -3 \]
\,  \exp\(- {\zeta \over R_0^{2n}} \) \cr & \qquad + 3 \, \[
\zeta^{{2 \over n}} \, \({ 1\over 2} - {2 \over n^2} \) \gamma\(1
- {2 \over n} , {\zeta \over R_0^{2n}} \) + {1 \over 2} \,
\zeta^{{1 \over n}} \, \gamma\(1 - {1\over n} , {\zeta \over
R_0^{2n}} \) \] \cr & \qquad + {1\over 2 \, n} \, \zeta^{{2 \over
n}} \, \sum_{m=3}^{\infty} \, a_m \({ 6 \over n}\)^{m/2} \,
\gamma\({m\over2} - {2 \over n} , {\zeta \over R_0^{2n}} \) }}
The first line is the contribution from $r \in [0,R_0]$. The
second line is the contribution from the gradient term and the
quadratic term in the potential. The last line includes the
contribution from the potential. The limit $ n \to 2$ is taken so
as to isolate the divergence coming from the quadratic term in the
potential and the $r^2 \, \(\phi_{,r}\)^2$ contribution. Using
\gamlim\ it is simple to show that,
 \eqn\admtwo{\eqalign{
{1 \over 2 \pi^2} \,  \Mc &= - {R_0^4 \over 2} \, \[ \exp\( 2
{\sqrt{\zeta} \over R_0^2} \) + 2 \, \exp\( - {\sqrt{\zeta} \over
R_0^2} \)
 -3 \]
\,  \exp\(- {\zeta\over R_0^4} \) \cr & \qquad + 3 \, \[ \zeta +
{1 \over 2} \, \sqrt{\zeta} \, \gamma\({1 \over 2}, {\zeta \over
R_0^4} \) + {\zeta \over 12}  \, \sum_{m=3}^{\infty} \, a_m
(3)^{m/2} \, \gamma\({m\over2} - 1 , {\zeta \over R_0^4} \) \] \ .
}}
While the above formula is ungainly it is easy to perform a small
$A$ expansion and we obtain,
\eqn\masstwosmall{ \Mc = \pi^2 \, \(1 + {2 \over R_0^2} \) \, A^2
+ {\cal O}(A^3) = \pi^2 \,\phi_0^2 \, R_0^4 + \cdots }

Note that if we first set $n=2$ then the first term in the second
line of \adm\ can be set to zero. This will lead to the negative
energy quoted in \hhm, \ie, $\Mc \approx - \pi^2 \, A^2$.

It is also useful to consider the value of the mass for the case
of large fields $\phi_0 \gg 1$. In this case, the negative
contribution from the inner region is exponentially suppressed
(since $\zeta \sim \phi_0^2  \gg 1$). Moreover, it is possible to
sum the series in \admtwo, and we obtain
\eqn\masstwolarge{ \Mc = 10.2348 \, \phi_0^2 \, R_0^4 + \sqrt{3}
\, \pi^{5/2} \, \po R_0^2 + \CO(e^{-\phi_0^2}) }

Another interesting limit to consider is $n \to \infty$ with $\po$
held fixed. This limit results in the field profile being a step
function. In this case from \adm, we obtain,
\eqn\stepmassa{ \Mc = 3 \, \pi^2 \, R_0^2 \, (1+R_0^2) \ , }
which is independent of $\po$!
\subsec{Generalization to other dimensions}

In AdS$_d$ one can parametrize a spherically symmetric Euclidean
slice $\Sigma_{d-1}$ by an appropriate generalization \sigmamet;
\eqn\sigmametd{ ds^2_{\Sigma} = \(1 - {m(r) \over r^{d-3} } + r^2
\)^{-1} dr^2 + r^2  \, d\Omega_{d-2}^2 \ . }
The analog of the constraint equation \constraints\ equates the
Ricci scalar of $d-1$ dimensional metric \sigmametd\ to the scalar
stress tensor, resulting in
\eqn\dmdiffeq{ m_{,r} + {1 \over d-2} \, r \, \(\phi_{,r}\)^2 \, m
= {2 \over d-2} \,  r^{d-2} \,
\[V(\phi) + {(d-1)(d-2) \over 2} + {1\over 2} (1+r^2)\,
\(\phi_{,r}\)^2
\] \ .
}
The mass of the configuration can be obtained as before by
integrating \dmdiffeq. The ADM mass for any asymptotically AdS$_d$
configuration is given as
\eqn\dadmm{ \Mc =  {1\over \Gamma\({d-1 \over 2} \) } \,(d-2) \,
\pi^{{d-1 \over 2}} \lim_{r \to \infty} \; m(r) }
%

\subsec{Scalar hair black holes}

Scalar hair black holes are solutions to the lagrangian
\HHMaction, with a non-trivial scalar field profile outside the
black hole horizon. We are interested in static spherically
symmetric black holes since we imagine that the scalar
configurations considered in Section 4, will ultimately settle
down into a static configuration. Hence we choose the following
metric ansatz:
\eqn\scalarhbh{ ds^2 = - e^{-2 \delta(r)} \, f(r) \, dt^2 +
{1\over f(r) } \, dr^2 + r^2 \, d \Omega_3^2 }
The equations of motion following from \HHMaction\ are:
\eqn\scbheqns{\eqalign{ \delta'(r)  &= - {1 \over 3} \, r \,
\phi'(r)^2 \cr r \, f'(r) - 2 + 2 f(r)  & = - {1 \over 3} \, r^2
\, \( f(r) \, \phi'(r)^2 + 2 \, V(\phi) \) \cr f(r) \, \( \phi'(r)
+ r \, \phi''(r) \) & = r \, {\partial V \over
\partial
\phi} + {2 \over 3} \, \phi'(r) \, (r^2 \, V(\phi) - 3) }}
To evaluate the black hole mass, it is simpler to parametrize the
function $f(r)$ in terms of the mass function $m(r)$ as
\eqn\fm{ f(r) = 1 + r^2 - {m(r) \over 3 \, \pi^2 \, r^2} }
It is simple to integrate for $\delta(r)$ and $m(r)$ in terms of
the scalar profile $\phi(r)$. Since we are looking at a static
spherically symmetric system, we find, not surprisingly, that
$m(r)$ is given in terms of $\phi(r)$ by the now familiar equation
\mdiffeq.

In order to calculate the mass of a scalar hair black hole of
horizon size $\rs$, we have to solve the equations \scbheqns\
numerically. However, one can get a reasonable estimate of the
mass in the weak field limit analytically. As explained in the
text, we take the linearized solutions for the scalar field in AdS
background and impose the appropriate boundary conditions. The
correct profile in the weak field limit is as given in \bhphia.

The mass of the black hole can now be computed using the formula
\massfn. We have two contributions: The contribution to the mass
from within the horizon can be reasonably approximated by the mass
of an AdS-Schwarzschild black hole of the same size. For the
region outside the horizon we take the profile to be as in
\bhphia. Moreover we can drop the exponential term in \massfn,
\ie, set $F(x) =1$, since $\phi$ is small. We therefore have:
\eqn\scalarbhmass{ M_{SBH} = 3 \pi^2 \, \rs^2 \, (\rs^2 + 1 ) + 2
\pi^2  \, \int_{\rs}^{R_1} \,dx \,
 x^3 \,\[V(\phi) + 6 + {1\over 2} (1 + x^2)\, (\phi_{,x})^2 \] \ ,
}
which integrates to
\eqn\appbhcut{ { M_{SBH} \over 2 \, \pi^2 } =
 {3 \over 2} \,  \rs^2 \, (1
+ \rs^2)  +\a_{bh}^2 \,
\[- \( {\ln (e \,R_1/\rs) \over \ln R_0 } \)^2 + { \ln R_1\over 2 \,
(\ln R_0)^2} \] \ .
 }
In deriving \appbhcut, we work in the regime $R_1 \gg R_0 \gg \rs
\gg 1$ and drop subleading terms like $1/\rs$ that show up upon
integration. The mass for the configuration \cutconf\ has been
evaluated in \masscut.

\subsec{Strong fields in finite radius cut-off AdS}

From Appendix A.2.\ we learn that for the finite radius cut-off
AdS space there is no parameter regime for weak fields with the
scalar profile as in \cutconf\ where black holes formation is
prohibited on simple kinematic grounds. One basic problem was that
in the weak field regime the singularity forms too late, which
means that the black hole that needs to form can be quite small
(see Appendix B.2). We learn there that for strong fields, the
singularity will form very quickly and in principle by going to
the non-linear regime $\rs \sim R_0$. This by itself is not good
enough as evidenced in \masstwolarge\ for non cut-off AdS
scenario. This is because the configurations we have been
considering so far are not the optimal ones. As discussed in the
text it is possible to set-up a minimization problem with
appropriate boundary conditions.

Since the weak field regime is not well suited to the formation of
naked singularities, we would like to examine the strong field
situation closely. In this case from the discussion of FRW
cosmologies with strong fields we learn that the putative horizon
radius is a finite fraction of the radius of the homogeneous
region; $\rs = \ss \, R_0$, with $\ss =0.577$ in five dimensions
\rsc\ ({\it cf.}, Appendix B.5).

Suppose the value of the scalar field at the horizon is $\phi(\rs)
= \ps \gg 1$. Consider the following field redefinitions: $
\psi(r) = \phi(r) - \ps$ and $m(r) = e^{{2 \over \sqrt{3}} \ps }\,
\mu(r) $. Since the scalar potential is a sum of two exponentials,
it takes a particularly simple form for large $\ps$; $V(\phi)
\approx -2\,  e^{{2 \over \sqrt{3}} \ps} \, \exp({2 \over
\sqrt{3}} \psi)$.

The equations of motion for $m(r)$ and $\phi(r)$, \scbheqns,
become very simple in terms of the redefined fields;
\eqn\scbheqnsa{ \mu'(r) + {r\over 3} \, \psi'(r)^2 \, \mu(r) = 2
\, \pi^2 \, r^3 \, V(\psi) + \CO(e^{-{2 \over \sqrt{3}} \ps}) }
\eqn\scbheqnsb{ -{1\over r^2} \, \mu(r)  \, \( \psi'(r)+ r \,
\psi''(r) \) = 2 \, \pi^2 \, \( \sqrt{3} \, r +  \psi'(r) \,r^2 \)
\, V(\psi) +
 \CO(e^{-{2 \over \sqrt{3}} \ps })
}
Here $V(\psi) = -2 \exp({2 \over \sqrt{3}} \psi) $ and we made use
of the fact that $V_{,\psi} = {2 \over \sqrt{3}} \, V_{\psi}$.
Ignoring all the terms containing the factor $\CO(e^{-{2 \over
\sqrt{3}} \ps })$ for the moment, we see that there is an
attractor solution to \scbheqnsb,
\eqn\attractor{ \psi_*(r) = \sqrt{3} \, \log\( {\rs \over r} \) \
. }
We picked the constant of integration to set $\psi(\rs) = 0$ as
required. This however cannot be the full story; the profile for
the field $\psi(r)$ cannot decay logarithmically in $r$, for if it
were, the geometry would not be asymptotically AdS. When does this
profile for $\psi(r)$ break down? As can be seen from the
equations it will happen when the terms multiplied by $ \CO(e^{-{2
\over \sqrt{3}} \ps })$ start to become significant. This will
happen at a length scale $R_* \sim (e^{{1 \over \sqrt{3}} \ps })$.
It is easy to engineer $\rs \ll R_*$ so that there is a large
region where we an use \attractor.   We restrict ourselves to
regimes where the $r > R_*$ region is negligible.

With $\rs \ll R_*$ and $\rs = \ss \, R_0$, it is now very easy to
estimate the mass difference between the black hole and the
configuration \nlcutconf. We basically need to solve for $\mu(r)$
from \scbheqnsa, and in particular need the solution only in the
range $r \in (\rs, R_0)$.
 Using \attractor, we obtain
\eqn\muinteg{ {d \over dr } \(r \, \mu(r) \) = - 4 \, \pi^2  \,
\rs^2 \, r^2 \ . }
For the black hole the boundary condition is $\mu_{sbh}(\rs) = 0$.
This follows because the black hole mass is independent of $\po$
and in the scaling regime the usual factor of $3 \, \pi^2 \, \rs^2
\, (1+\rs^2)$ is exponentially suppressed. This implies,
\eqn\nlsbhmass{ M_{SBH} =  - {4 \over 3} \,\pi^2 \, \ss^2 \, R_0^4
\, (1 - \ss^3) \; \;   e^{{2 \over \sqrt{3}} \ps } \, e^{-{1 \over
3} \,  \int_{R_0}^{R_1} \, x \, \psi_{sbh}'(x)^2} \ . }
On the other hand, for the configuration \nlcutconf\ we have
$\mu_{{\rm config}}(R_0) = -  \pi^2 \, R_0^4 \, e^{2  \,
\psi_*(R_0)/\sqrt{3}} =  -\pi^2 \, \rs^2 \, R_0^2$. So we get
\eqn\nlsbhmass{ \Mc =  - {1 \over 3} \, \pi^2
 \, \ss^2 \, R_0^4 \;\; e^{{2 \over
\sqrt{3}} \ps } \, e^{-{1 \over 3} \,  \int_{R_0}^{R_1} \, x \,
\psi_{sbh}'(x)^2}  \ . }
One of  prefactors in the mass formulae can be traced to the
scaling, while the other is related to the fact that as we go out
to larger length scales the contribution to the mass from the
inner regions gets exponentially suppressed as in \massfn.
 The mass difference is now
\eqn\nlmassdiffapp{ M_{SBH} - \Mc =   {\pi^2 \over 3} \, \ss^2 \,
R_0^4 \, (4 \, \ss^3 - 1) \;\;
 \,  e^{{2 \over
\sqrt{3}} \ps } \, e^{-{1 \over 3} \,  \int_{R_0}^{R_1} \, x \,
\psi_{sbh}'(x)^2} \;\; < \;\; 0 \ , }
since $\ss = 0.577$.

\appendix{B}{Field dynamics in $d$ dimensions}

We want to ask whether naked singularities are more likely to form
in dimensions other than 5, and examine (for particular dimensions
of interest) whether there might be numerical ``windows of
opportunity'' for initial conditions that would lead to naked
singularity. This involves 1) verifying that a homogeneous scalar
field collapses to a singularity, 2) estimating the mass of a
sufficiently large black hole which would cloak the singularity,
and 3) estimating the mass of the starting configuration (as
already discussed in Appendix A).

In Appendix B.1 we study the FRW cosmology in $d$ dimensions and
show that the cosmologies big crunch in all dimensions. In B.2 we
make an analytic estimate of the size of the singularity in the
weak field regime. In B.3 we exhibit a scaling argument valid for
the 5d supergravity potential in the strong field regime. In B.4
we study the size of the singularity numerically, using the same
physical prescription as in B.2. This allows us to get a better
estimate for the numerical coefficients in the weak field limit
and also understand the strong field behaviour. Finally, in B.5 we
present a more sophisticated argument for determining the size of
the singularity and give our numerical results. While the main
focus of this paper is on cosmic censorship in the AdS/CFT
context, for completeness, in B.6 we discuss the corresponding
estimates in the context of the double-well potential of \hhmgr.

\subsec{Evolution toward singularity}

The first step is to model the evolution of a homogeneous scalar
field in hyperbolic slices of AdS. This is an accurate description
of the dynamics in the domain of dependence of the homogeneous
region $r \le R_0$ for the configurations considered in the body
of the paper. The metric in $d$ dimensions can be written as
\eqn\frwmetd{ ds^2 = -d\, T^2 + a(T)^2 \, \( {dR^2 \over 1+R^2 } +
R^2 \, d\Omega_{d-2}^2 \) }
where the scale factor $a(T)$ and the scalar field $\phi(T)$ are
to be determined from the Einstein's equations  $G_{\mu \nu} =
T_{\mu \nu}$ (having set $8 \pi G_N \equiv 1$).

From \frwmetd\ we can readily find the Einstein tensor:
\eqn\Einstd{ G_{TT} = {1 \over 2} (d-1) (d-2) \, { \ad^2 - 1 \over
a^2 } \ , \qquad G_{RR} = - {(d-2) \over 2} \,
  { 2 \, a \, \add + (d-3) (\ad^2 - 1) \over 1 + R^2 }
}
(As an aside, note that the Ricci scalar is
\eqn\Riccid{ \CR = (d-1) \,  { 2 \, a \, \add + (d-2) (\ad^2 - 1)
\over a^2 }}
This indicates a curvature singularity as $a \to 0$, unless $ a(T)
\sim \cos T $ as for pure AdS, where $a = 0$ corresponds merely to
a coordinate singularity.  But as we will see below, for
nontrivial scalar field, $a$ cannot be a simple trigonometric
function.)
We can also find the stress tensor in terms of the (as yet
unknown) potential $V(\phi)$, by evaluating $T_{\mu \nu} =
\del_{\mu} \phi \, \del_{\nu} \phi - {1 \over 2} \, g_{\mu \nu} \,
\[ \del_{\mu} \phi \, \del^{\mu} \phi + 2 V \] $; in the present
context,
\eqn\stressd{ T_{TT} = {1 \over 2} \, \pd^2 + V \ , \qquad T_{RR}
= {a^2 \over 1+ R^2} \, \( {1 \over 2} \, \pd^2 - V \) }
Putting \Einstd\ and \stressd\ together, we obtain
\eqn\evocnsd{ \ad^2 - {2 \over (d-1)(d-2)} \, a^2 \, \( {1\over 2}
\, \pd^2 + V \) = 1 }
\eqn\evolvead{ {\add \over a} - {2 \over (d-1)(d-2)} \,
            \(  V - {(d-2) \over 2} \, \pd^2 \) =0
}
Finally, from the stress tensor conservation $\nabla^{\mu} T_{\mu
\nu} = 0$, we can derive the (linearly dependent) third equation
\eqn\evolvephid{ \pdd + (d-1) \, {\ad \over a} \, \pd  + V_{,
\phi} = 0 }

Evolving these equations is difficult to do analytically, but it
is relatively straightforward numerically---once we know the full
$V(\phi)$. Finding $V(\phi)$ suited for the present purposes is
however often prohibitively difficult. Nevertheless, there are 2
regimes where we can make progress: Firstly, for small fields, we
can use the quadratic expansion of $V$ for a BF-saturating field;
in particular, the constant term is fixed by the AdS radius (which
we'll set $=1$) and the quadratic term by $m_{BF}^2 = - { (d-1)^2
\over 4 }$. Thus, $V(\phi) = - {(d-1) (d-2) \over 2 } -  { (d-1)^2
\over 8 } \phi^2 + \CO(\phi^3) $. Numerically evolving $\phi(T)$
and $a(T)$ in this potential leads to a curvature singularity at
$T < {\pi \over 2}$ (though of course in the process $\phi \to
\infty$ and the approximation to $V$ is no longer valid).

Secondly, however, we can study the near-singularity regime, where
the potential is negligible. Neglecting the third term in
\evolvephid\ we can solve for $\pd(T)$ in terms of $a(T)$:
\eqn\phida{ \pd(T) \approx  { c \over a(T)^{d-1} } }
where $c$ is a constant. Substituting \phida\ back into \evocnsd\
(and ignoring $V$ and the $1$, since they are both negligible near
the singularity), we obtain a first order differential equation
for $a(T)$: $\ad^2 = {c^2 \over (d-1)(d-2)} \, {1 \over
a^{2d-4}}$. Taking the (negative branch) square root and
integrating, we have
\eqn\adT{ a(T) \approx \[ c \, \sqrt{{d-1 \over d-2}} \, (T_s - T)
\]^{{1 \over d-1}} }
where $T_s$ corresponds to the time of the singularity, $a(T_s) =
0$. Finally, using \adT\ in \phida\ and integrating yields
$\phi(T)$:
\eqn\phiT{ \phi(T) \approx \sqrt{{d-2 \over d-1}} \, \ln(T_s - T)
+ {\rm const} }
This verifies that there is a singularity in all dimensions.

\subsec{Black hole estimates}

Let us now turn to the second step, namely estimating the mass of
the black hole required to cloak the singularity. This involves a
bit more detailed analysis, since it depends on the behaviour of
the null geodesics in the dynamical background. We will use the
same procedure as before, and estimate at what time does the pure
AdS metric cease to approximate the actual geometry.  We will then
use that as the cutoff for the ingoing radial null geodesic
starting at $T=0$ at radius $R_0$; namely, we approximate the
black hole size $\rpa$ by the proper radius reached by this null
geodesic at the cut-off time $\Tco$, \cf, Fig.4. In Appendix B.4
we will contrast these analytical estimates with numerical
calculations in the FRW geometry.

Since we are starting with a time symmetric configuration at
$T=0$, at early times $\pd(T) \approx 0$, so that from \evocnsd\
we have $\ad^2 - {2 \, V \over (d-1)(d-2)} \, a^2 =1 $. Denoting
$H^2 =  - {2 \over (d-1)(d-2)} \, V(\phi_0)$ as before, we find
the early time solution
\eqn\earlya{ a(T) \approx {\cos(H \, T) \over H} }
(Note that $H \ga 1$, so that the singularity must occur at $T_s
\la {\pi \over 2}$.  That is, for pure AdS, there is a coordinate
singularity when $a$ vanishes at $T_s = {\pi \over 2H}$; and
whenever there is a nontrivial field, $a(T) \to 0$ even faster, as
evident from \evocnsd. But in that case, the singularity is a
genuine curvature singularity since $\phi \to \infty$ in addition
to $a \to 0$.) From \evocnsd\ we see that the scale factor starts
deviating from this form when
\eqn\anewform{ a^2 \, \pd^2 \sim (d-1)(d-2)}
  What is $\phi$?  Initially, $\phi$ grows
exponentially as it rolls down the quadratic potential, $\phi(T)
\approx \po \, \cosh({d-1 \over 2} \, T)$ where $\po=\phi(0)$ is
the initial value of the field in the internal region; but at
relatively early times $T_0 \approx {1 \over \sqrt{d-1}}$, when
$V_{, \phi} < {\ad \over a} \, \pd$, $\phi(T)$ changes form and
starts behaving as in \phida. Matching at $T_0$, we determine the
constant $c \approx \eta \, \po$, where $\eta \approx 0.44$ for
$d=3$ and $1.41$ for $d=5$. Substituting this back into \anewform,
we see that the scale factor changes form when $a \approx \[
(d-1)(d-2) \]^{-{1 \over 2d-4}} \, c^{{1 \over d-2}}$. Matching to
the initial form of $a$, we find that the metric no longer
approximates pure AdS at time $\Tco$, given by
\eqn\epsd{ \e \equiv {\pi \over 2 H} - \Tco = \upsilon \, \po^{{1
\over d-2}} }
For $d=3$, $\upsilon \approx 0.31$, whereas for $d=5$, $\upsilon
\approx 0.74$. (Note that at this point, the field is still small:
$\phi(\Tco) \approx \po^{{1\over d-1}} \ll 1$ for $\po \ll 1$; so
the quadratic approximation of the potential should be reasonable,
and we can easily verify that at $\Tco$, $V \ll \pd^2$.)

\ifig\figRestim{``Causal'' diagram in FRW coordinates (figure not
to scale) for obtaining estimates of the horizon size.}
{\epsfxsize=7.5cm \epsfysize=3.8cm \epsfbox{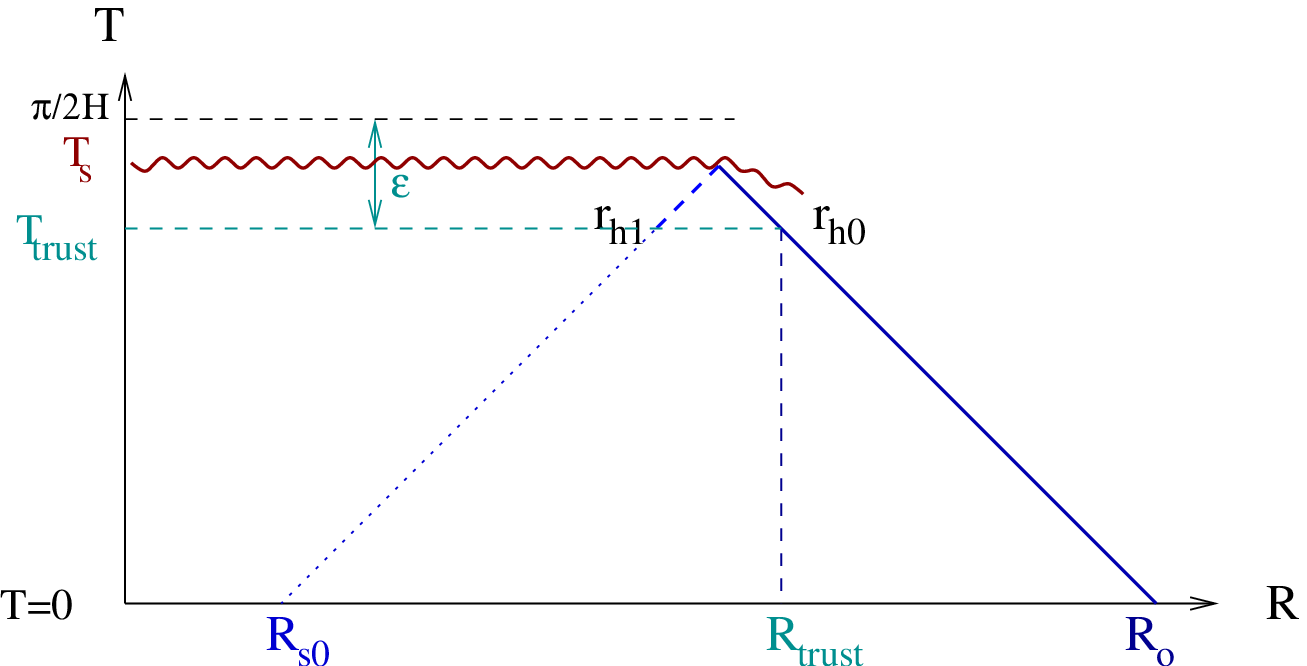}}

We now estimate the black hole size by simply taking it to be the
proper radius $\rpa \approx r(\Tco) = R(\Tco) \, a(\Tco)$,
determined by ray-tracing back the ingoing radial null geodesic
starting from $R_*$ at $T=0$ (\cf\ \figRestim). Solving for the
radial null geodesics in the geometry \frwmetd\ with \earlya, we
obtain\foot{This formula is valid even in the strong field regime
as long as $T \le T_{trust}$. In the weak field regime $H \approx
1 + {1\over 6} \po^2 \sim 1$, and so
 $R_* \sim R_0$. In the strong field regime
we have to take into account the fact that proper radius of the
spheres are shrunk by a factor of $H$; since $R_0$ is proper size
of the homogeneous region in unit sized AdS geometry, we should
take $R_0 = H \, R_*$.}
\eqn\RTAdS{ R(T) = { R_* \, (R_* + \sqrt{R_*^2 + 1}) \, \cos^2(H
\, T) \pm \sin (H \, T) \over   (R_* + \sqrt{R_*^2 + 1}) \, (1 +
\sin (H \, T) ) \, \cos(H \, T)} }
where the $\pm$ denotes outgoing and ingoing geodesics,
respectively. Note that $R(T)$ diverges when $\cos(H T) = 0$;
however, the proper radius $r(T) = R(T) \, {\cos(H \, T) \over H}$
stays finite there. At $T=\Tco = {\pi \over 2 \, H} - \e$, the
scale factor is simply $a(\Tco) = {1 \over H } \, \cos({\pi \over
2} - \e \, H) \approx \e$ and \RTAdS\ also simplifies. To the
first subleading terms in $\e^2$, $\po^2$, and $1/R_0^2$, $\Rco =
R(\Tco)$ becomes
\eqn\Rtrust{
 \Rco \approx {-1 + 2 \, R_0^2 \, \e^2 \,
     (1+{5 \over 8 \, R_0^2}+{1\over 2} {d-1 \over d-2} \, \po^2)
\over
 4 \, R_0 \, \e }
 \, \[ 1 - {1 \over 4 \, R_0^2} - {1\over 2} {d-1 \over d-2} \, \po^2
\]
}
Note that in order to maintain $\Rco$ being positive, so that the
domain of dependence of the internal region includes the
singularity, we must have $ 2 \, R_0^2 \, \e^2 > 1$.  Using \epsd,
we must satisfy the following constraint on $\po$ and $R_0$:
\eqn\Rphbound{
 R_0 \, \po^{{1 \over d-2}} > {1 \over \upsilon \sqrt{2}} \sim 1 \ . }
Kepping all the coefficients in the above estimate, the RHS is
0.96 for $d=5$ and 2.28 for $d=3$. Without this condition, we are
not guaranteed that the initial field configuration evolves to a
singularity.

Note that \hhm\ have a much more conservative procedure to
estimate the horizon radius:  They ray-trace the null geodesic
from $R_0$ up to the singularity and then back to the initial data
slice, to obtain $\rsz$, as sketched in \figRestim.  (Actually, to
be able to ray-trace in pure AdS geometry, they bounce off from
$T=\Tco$ rather than $T=T_s$.)  The logic is that by the area
theorem for black holes, a real horizon, which on the initial data
slice must be larger than $\rsz$ in order to cloak the
singularity, could only grow in time. So $\rsz$ gives a {\it lower
bound} on the horizon radius and correspondingly the black hole
mass $\Mbh$. To show that naked singularities must be produced, we
want to show that the lower bound on $\Mbh$ is still larger than
$\Mc$, so we want to make this lower bound of $\Mbh$ as large as
possible. As one can easily check, in AdS geometry, the proper
radius $r(T) = a(T) \, R(T)$ grows along the outgoing geodesics.
This means that it is profitable to estimate the horizon radius by
the radius near\foot{ One has to be a bit careful, because in the
near-singularity regime \adT, $r(T) \to 0$ as $T \to T_s$ along
 outgoing radial geodesics which really terminate at the singularity
in the domain of dependence of the inner region.}
 the singularity rather than at the initial data slice,
\ie, by $\rpt$ in \figRestim.

Note however that above, we estimate the black hole size by $\rpa$
rather than $\rpt$ (\cf\ \figRestim). To check how much error we
are making in estimating $\rpa$ based on terminating ingoing null
geodesic at $T=\Tco$, rather than continuing all the way to the
singularity, we can estimate what the proper radius $\rpt$ would
be if we ray-traced the ingoing null geodesic from $\rpa$ up to
$T_s$ in the near-singularity geometry, bounced from the
singularity, and ray-traced backwards along the outgoing null
geodesic back to $\Tco$. We then obtain $\rpt = a(\Rco) \, \sinh
\[ \sinh^{-1} \Rco - 0.2 \]$, where the numerical factor 0.2 is
only very weakly dependent on $d$. Thus, for $\Rco > \CO(1)$, we
can check that we are not making a very large error: $\rpt \sim
0.82 \, \rpa$.

Now, suppose $ R_0 \, \po^{{1 \over d-2}} \gg 1$, so that the
singularity forms in a large region, $\Rco \sim \half \, \e \,
R_0$. Then the horizon radius which would cloak this singularity
would have to be at least
\eqn\rhor{ \rpa \sim a(\Tco) \, \Rco \approx {\upsilon^2 \over 2}
\, R_0 \, \po^{{2 \over d-2}} \ .  }
In five dimensions we find $ \rpa = 0.27 \, R_0 \, \po^{2/3}$.
Note that since this corresponds to the proper radius, which
includes the shrinking scale factor, this quantity may or may not
be large, depending on $\po$. In the text we quote $\rpa$ in
equation \rsa, as a lower bound on the size of the black hole.

We should point out that the  order one coefficient in the above
formula can be accurately  determined numerically, by studying the
FRW equations \evocnsd, \evolvephid, see Fig 6. Our numerical
studies in five dimensions shows that  $\rpa = 0.37 \, R_0 \,
\po^{2/3}$. The correction comes from the inaccuracy in  our
ability to accurately pin down the matching point $\Tco$  and the
time to the singularity $T_{s}$ (we have been using the upper
bound ${\pi \over 2\,  H}$).

The mass of the Schw-AdS$_d$ black hole is given by $\Mbh = (d-2)
\, \om_d \, \rpa^{d-3} \, (\rpa^2 + 1)$ where $\om_d \equiv
\pi^{{d-1 \over 2}} / \Gamma \( {d-1 \over 2} \) $). If $\rpa \gg
1$, then we can estimate the mass of the large AdS black hole by
\eqn\Mbhlarge{ \Mbh = \om_d \, {\varsigma} \, R_0^{d-1} \,
\po^{{2d-2 \over d-2}}}
where  $\varsigma \approx 0.0023$ for $d=3$ and
 $\varsigma \approx 0.017$ for $d=5$.

For the configuration mass given approximately by $\Mc \sim \om_d
\,  R_0^{d-1} \, \po^2$, we have (in the large black hole
approximation)
\eqn\mularge{ \mu \sim \varsigma \, \po^{{2 \over d-2}} \ll 1 \ .
}
This means that we are kinematically likely to produce black
holes, \ie, the energy of the starting configuration greatly
exceeds that of the requisite black hole. As we see, $\mu$ gets
larger (and therefore slightly more favorable to produce naked
singularities) for larger $\po$ and in higher dimensions; but
since $\varsigma \ll 1$, unfortunately there is no numerical
window of  opportunity wherein we could have $\po < 1$ while $\mu
> 1$. In \mularge, $\mu > 1$ would require $\po > \CO(100)$ which
is well outside the linearized regime where \mularge\ was valid.


\subsec{FRW cosmology with strong fields}

In Appendix B.1. we have modeled the dynamics in the domain of
dependence of the homogeneous central region of all of our
configurations in terms of FRW cosmology with hyperbolic spatial
sections. The details of the dynamics depend on the dimension and
on the precise form of the scalar potential $V(\phi)$. We are
interested in the asymptotic large $\phi$ behaviour for the
exponential supergravity potential \HHMpot. For $\phi \gg 1$ we
can ignore the second term in \HHMpot\ and so we will consider a
generic negative definite exponential potential $V(\phi) = - v \,
\exp(2 \, k \, \phi)$. The equations governing the dynamics are
any two of \evocnsd, \evolvead, \evolvephid, and  we have time
symmetric boundary conditions along with $\phi(0) = \phi_0 \gg 1$.

Introduce a new field $\psi = \phi - \po$. In terms of $\psi$ we
see that $V(\psi) = -v\,  e^{2 k \, \po} \, \exp(2 \, k \, \psi)$.
Since the potential rescales with the field redefinition, it is
tempting to scale the factor $ e^{k \, \po}$ out of the problem.
In fact, by rescaling the time coordinate $T = e^{- k \, \po} \,
U$ and the scale factor $b(U) = e^{k \, \po} \, a(U)$ we obtain
the equations:
\eqn\revocnsd{ \bd^2 - {2 \over (d-1)(d-2)} \, b^2 \, \( {1\over
2} \, \psd^2 - v \, e^{ 2 \, k \, \psi}  \) = 1 }
\eqn\revolvebd{ {\bdd \over b} + {2 \over (d-1)(d-2)} \,
            \( v \,  e^{ 2 \, k \, \psi} +
{(d-2) \over 2} \, \psd^2 \) =0 }
\eqn\revolvepsid{ \psdd + (d-1) \, {\bd \over b} \, \psd -2 \, v
\, k \, e^{ 2 \, k \, \psi} = 0
 }
We use the overdot to denote derivatives with respect to $U$,
$\dot{} \equiv {d \over dU}$. This rescaling has the effect that
the scale $e^{k \, \phi_0}$ drops out of the dynamics. In the
resulting rescaled problem the time to singularity will be
determined solely by the number $\tilde{H}^2 = {2 \, v \over
(d-1)(d-2)}$. The equations \revocnsd, \revolvebd, \revolvepsid,
will collapse to a singularity $ b(U) \to 0$ in a finite time
$U_s$, which is a finite fraction of the rescaled AdS time $\pi/2
\, \tilde{H}$. Hence we see that $T_s \, e^{k \, \po}$ is finite;
the homogeneous region exists for a finite fraction of the
original AdS time $\pi / 2 H = \pi \, e^{k \, \phi_0}/2 \,
\tilde{H}$ before the big crunch singularity.

\subsec{Numerical estimates of black hole size}

In Appendix B.2 we discussed the size of the black hole required
to cloak the singularity. Most of our estimates there were based
on $\po \ll 1$, and we derived the formula $\rpa \sim R_0 \,
\po^{2/3}$ in this regime.   We will improve on this below. It is
clear that as we increase $\po$ the singularity will happen on
much shorter timescales, since the backreaction on the geometry is
going to be more important. One might naively think that in the
regime $\po \gg 1$ the singularity will encompass all of the
homogeneous region and in particular estimate $\rpa \sim R_0$.

To get a reliable estimate on the size of the black hole necessary
to cloak the singularity, we solve FRW equations \evocnsd,
\evolvead, and \evolvephid\ numerically. With an explicit
numerical solution to the scale factor $a(T)$, we can study the
property of null geodesics in the resulting big crunch geometry.

\ifig\figstrrplus{Numerical estimates for $\rmx(\phi_0)$. (a)
$\rmx(\phi_0)/R_0$ evaluated from integrating the geodesic
equations. (b) Proper radius as a function of time in the rescaled
geometry. The maximum value is attained at $ T = T_{max}$. }
{\epsfxsize=12cm \epsfysize=3.8cm \epsfbox{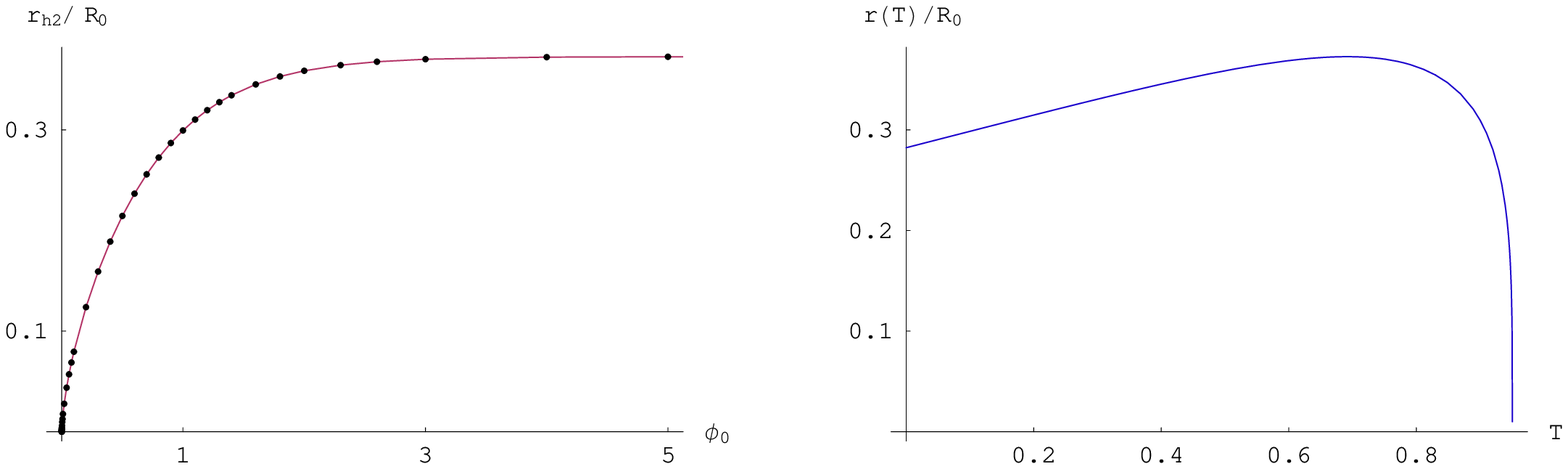}}

In particular we can obtain the behavior of $\rpa(\phi_0)$.  As
indicated in \figRestim, we start from a point on the initial data
slice where the spheres have proper size $R_0$ and consider a
radially ingoing null geodesic. At some point very close to the
singularity we cut-off the geometry and from the end-point of the
geodesic from $R_0$ onto this cut-off surface we take a radially
ingoing null geodesic. This would intersect the initial data slice
at $\rsz$. However, the proper size of the spheres will peak at
some intermediate time $T_{max}$. Call this radius $\rmx$; this
will be the lower bound on the size of the black hole. We use the
numerical solution for $a(T)$ to estimate $\rmx$. In \figstrrplus\
we show the behavior of $\rmx(\po)/R_0$. We see that for large
$\po$, $\rmx$ is independent of $\po$ as expected. Moreover, we
find that for small $\po$,
\eqn\rmax{ \rmx(\po) = 0.37 \, R_0 \, \po^{2/3}}
As mentioned in Appendix B.2, the numerical evaluation of $\rmx$
differs from our weak field estimate $\rpa$ only in the numerical
factor, which is attributable to a more accurate determination of
the point where the geometry deviates significantly from pure AdS.

An extremely efficient way to estimate the asymptotic value of
$\rmx(\po)$ as $\po \to \infty$ is to use the scaling argument of
Appendix B.3. We basically just have to solve the rescaled
equations, say \revocnsd, and \revolvepsid\ to find $b(U)$.  Armed
with the knowledge of $b(U)$ we can trace the null geodesics and
see where the maximum occurs (see \figstrrplus).  We find
\eqn\limrmax{ \rmx  \to  0.372 \, R_0 \ ,\qquad {\rm  for} \;\;\;
\po \gg 1 \ .}
%

\subsec{Refining the estimate of the black hole size}

The estimates for the black hole size \rmax, \limrmax, do not
constitute the tightest lower bound. In deriving them we assumed
that the singularity stretched out to the boundary of the domain
of dependence at time $T_s$. However, we now argue that it
necessarily has to be bigger and so obtain more stringent bounds.

In estimating $\rmx$ we used the essential fact that the
congruence of outgoing null geodesics starting from a point $\rsz$
on the initial data surface first expands and then contracts in
size, attaining a maximum at $T_{max}$. This implies that the
surface of constant FRW time, $T= T_{max}$, is a marginally
trapped surface. Since the region $T > T_{max}$ is a region of
trapped surfaces, any geodesic entering it must necessarily end at
the singularity.

\ifig\figrpest{``Causal'' diagram in FRW coordinates (figure not
to scale) with the new information to refine the size of the
singularity.} {\epsfxsize=7.5cm \epsfysize=3.8cm
\epsfbox{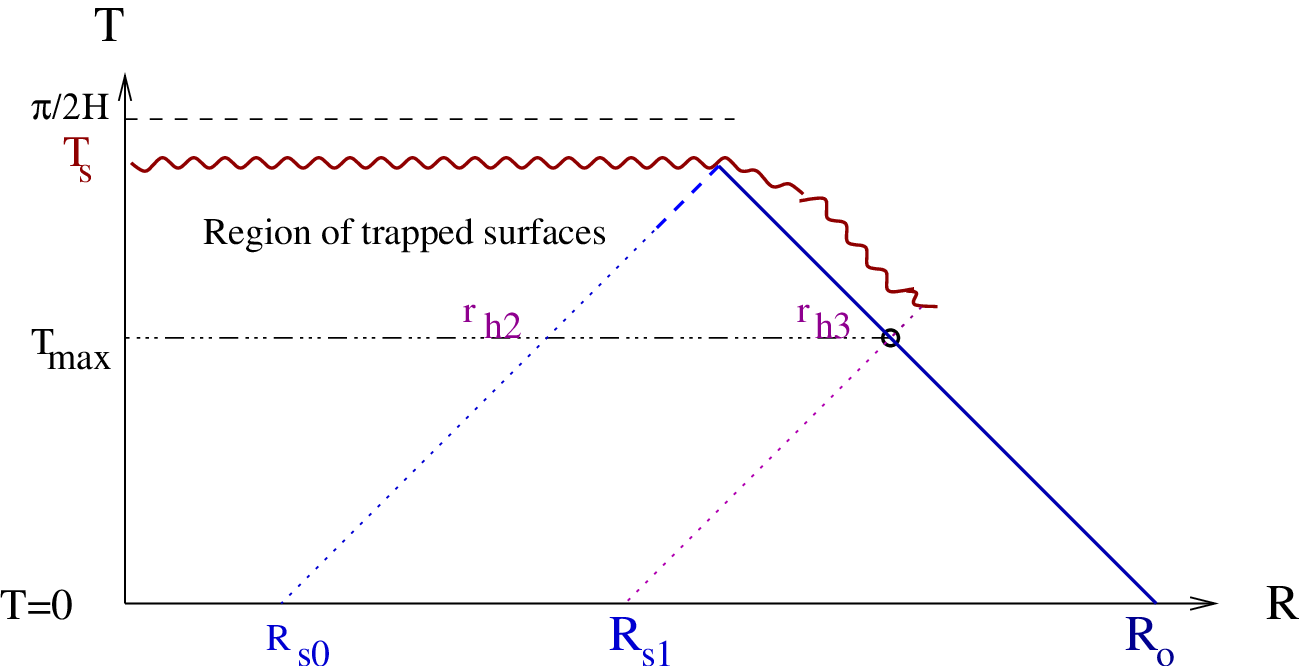}}

Hitherto, we picked the congruence starting from $\rsz$  that
meets up at the singularity with the ingoing null congruence
emanating from $R_0$. Now consider an outgoing null congruence
that starts at $\rsz + \delta R$. This congruence will initially
start out expanding. Can it expand for ever and reach the boundary
of AdS? We will argue that for sufficiently small $\delta R$ it
can not. First of all, the geodesics  will have to enter the
region of trapped surfaces $T > T_{max}$, and hence will start
contracting in this region. Moreover, it is also clear that this
congruence will exit the domain of dependence of the homogeneous
central region before meeting with the singularity, see \figrpest.
If these geodesics were to miss the singularity, then they will
end up at the AdS boundary. However, this will be a clear
violation of the Raychaudhuri equation for geodesic congruences,
which implies that a contracting congruence of null geodesics
cannot re-expand, so long as the matter contributing to the stress
tensor satisfies the null energy condition\foot{ For the
scalar-gravity action \HHMaction, the stress-tensor is given
before \stressd. Despite the negative-definite potential, the null
energy condition, $T_{\mu \nu} k^{\mu} k^{\nu} \ge 0$ $\forall$
null $k^{\mu}$ is satisfied, since the contraction with a null
vector projects out the potential contribution.}. For the
congruence in question, there is only one possible outcome -- it
must end up at the singularity once it enters the region of
trapped surfaces. This in particular implies that the singularity
must extend outside the domain of dependence of the initial data
surface. A similar argument was used by \hhm\ to argue that the
singularity should have a finite proper size.

\ifig\figrefrp{Numerical estimates for $\Rp(\phi_0)$.
\smallskip
\noindent (a) Comparing $\Rp(\po)/R_0$ (upper curve) to
$\rmx(\po)/R_0$ (lower curve).
\smallskip
\noindent (b) Weak field estimate of $\Rp(\po)$ and comparison to
the analytic fit \rpweak. } {\epsfxsize=12cm \epsfysize=3.8cm
\epsfbox{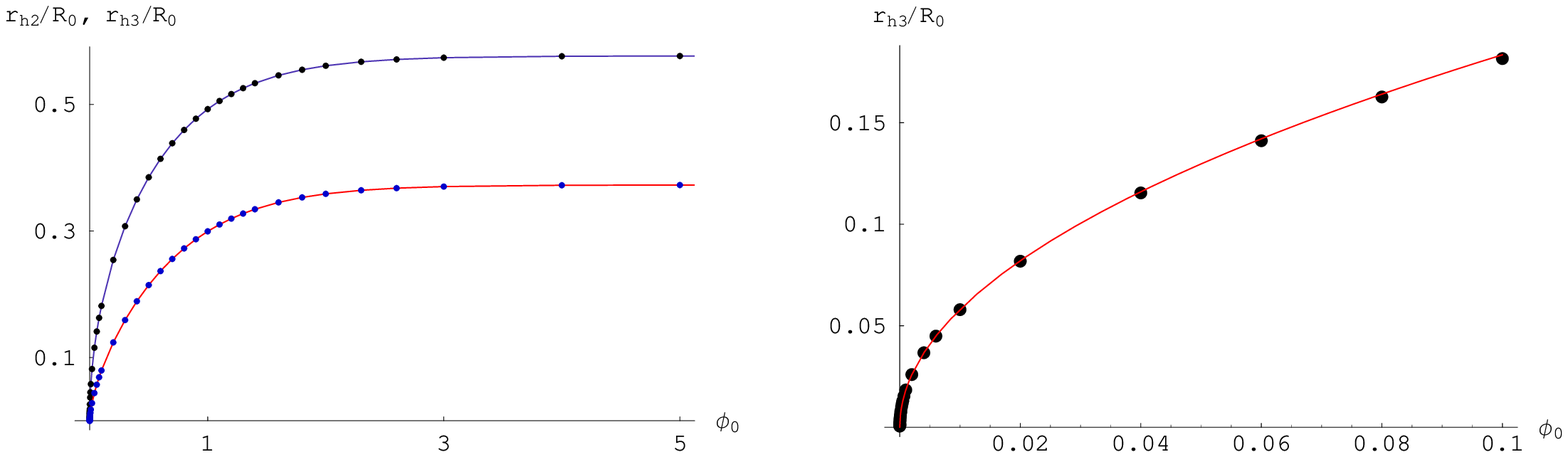}}

Having established that the singularity must stretch outside the
domain of dependence of the region $r \le R_0$, we would like to
obtain the best estimate for $\delta R$. On the constant FRW time
$T=T_{max}$ surface, the proper size of the geodesics is a
monotonically increasing function of $\delta R$. This implies that
the best estimate for the proper size is achieved by moving out to
the boundary of the domain of dependence. So we only need to know
the proper size of the spheres at the boundary of the domain of
dependence (and hence by definition on the ingoing congruence from
$R_0$) at $T_{max}$. From \figrefrp, we see that the outgoing null
congruence from $\rso$ is the one where the maximum proper size
$\Rp$ is reached at the boundary. For larger starting $R$, there
is no maximal size sphere along an outgoing congruence within the
domain of dependence of the homogeneous region, and we would have
to resort to full numerical simulation to make further refinements
to our estimate.

Using the numerical solution for the scale factor one can
ascertain $\Rp$. We show in \figrefrp\ the numerical estimates for
the same and compare it also to the previously derived results of
Appendix B.4. We find that in the weak field limit we have
\eqn\rpweak{ \Rp(\po) = 0.58 \, R_0 \, \po^{1/2} \ , \qquad {\rm
for} \;\;\; \po \ll 1 \ . }
This is parametrically larger than our previous estimate \rmax\ by
a factor of $\po^{-1/6}$. In the strong field limit we obtain
\eqn\rpstr{ \Rp \to \ss \, R_0 = 0.577 \, R_0 \ , \qquad {\rm for}
\;\;\; \po \gg 1 \ . }
This betters our previous estimate \limrmax\ by about 55\%, which
is a decent improvement, but sadly insufficient to rigorously rule
out black hole formation. It is likely that the singularity
stretches out a bit more than what we estimated above, but this is
not something we can estimate reliably within the domain of
dependence of the homogeneous region \ie, in the FRW cosmology. We
therefore make our estimates for black hole masses using the size
of the horizon to be given by $\Rp$. The values of $\Rp$ are
quoted as the lower bound for $\rs$ in the text in \rsb\ and \rsc,
for weak and strong fields, respectively.

These results can easily be generalized to other dimensions. In
order to obtain the strong field behaviour of $\Rp$,
 we need to know the exact form of the scalar potential.
In the weak field limit we can make progress using the quadratic
approximation to the scalar potential. For configurations
analogous to \HHMphi, the mass can be determined from \dadmm\ to
be $\Mc ={ (d-1) \over 4 } \,  \omega_d \, \po^2 \, R_0^{d-1}$.
The black hole mass, with horizon size $\Rp$ is of  course,
$M_{BH} = (d-2) \, \omega_d \, \Rp^{d-3} \, (\Rp^2 + 1) $. In
$d=7$, with $\po = 0.01$ we find $ \Rp = 0.148 \, R_0$, leading to
the parameter $\mu$, as defined in \mudef, being $\mu = 0.35$. For
smaller $\po$, say $\po = 0.001$, we find $\Rp = 0.0693 \, R_0$
implying $\mu = 0.37$. We see therefore that $\mu$ gets a little
bigger as the number of dimensions increases, but far too slowly
to make the case for singularity formation in higher dimensions.

\subsec{The double well potential scenario}

Let us now return to the double well potential scenario discussed
in \hhmgr.  As mentioned in the text, there is a gap in the
arguments for violating cosmic censorship, which we now briefly
discuss. The minimizing field configuration, which is guaranteed
to have energy $\sim R_0$, does not have exactly constant profile
in the interior region. As suggested by the numerical analysis of
\garfinkle\ and later explained by \hhmrev, this leads to the
actual size of the singularity being smaller than assumed in
\hhmgr, and in fact the black hole required to cloak this smaller
singularity may require less mass than that available in the
configuration.

Conversely, suppose we keep the field profile homogeneous in the
interior region.  One might hope this does not increase the mass
too much, whereas the size of the singularity is as given by  the
original estimates.  If that were the case, then as long as one
started with sufficiently large $R_0$, one could retain the
kinematic arguments for forming naked singularities.
Unfortunately, the mass does increase too much,  a point made
independently in \hhmrev . Here we will demonstrate this statement
numerically, since we can use essentially the same analysis as
above to bound the size of the singularity and thereby ascertain
whether or not the requisite black hole mass is sufficiently
large.

For small fields, we can use the quadratic approximation of the
potential. Consider therefore the dynamics of the scalar field in
the potential given by $V(\phi) = -3 + {1 \over 2} m^2 \, \phi^2$.
This generalizes the situation considered by \hhmgr\ by
introducing an arbitrary mass parameter $m$.  While the field
typically oscillates in the potential with the frequency given by
$m$ and then diverges logarithmically, the scale factor vanishes
in very similar manner as before. Repeating the numerical analysis
of the preceding section, we obtain (cf.\ \figrpest) $\rmx \sim
\po$ while $\Rp \approx c \, \po^{2/3}$, with $c \sim 0.2 \,
m^{4/3}$ for small $m$. Even for larger $m$ the coefficient
remains small; for example, for
 $m^2 =  100$, $c \approx 0.3$.
Using the estimate for the mass of the configuration, $\Mc \sim
m^2 \, \po^2 \, R_0^3$, we can easily calculate the parameter
$\mu$.
%
%
\midinsert
\centerline{%
\vbox{
  \offinterlineskip \tabskip=0pt
  \halign{\strut
          \vrule#&              %
          \hfil $ #~$ &\vrule#& %
          \hfil $\,#$ &         %
        ~ \hfil $#$ &\vrule#&   %
          \hfil $\,#$ &         %
        ~ \hfil $#$ &\vrule#&   %
          \hfil $\,#$ &         %
        ~ \hfil $#$ &\vrule#&   %
          \hfil $\,#$ &         %
        ~ \hfil $#$ &\vrule#    %
          \cr
     \noalign{\hrule}
     \noalign{\hrule}
 & \omit ~$\mu$
       &&  \omit \hfil $ \ \ \phi_0 = 0.0001$ \hfil & \omit \hfil 
                       $ \ \ \ \ \ \ \phi_0 = 0.1$ \hfil & \cr
     \noalign{\hrule}
    & \ m^2 = 100   && 0.00015   &  0.05  &   \cr
    & \ m^2 = 1   && 0.024  &  0.028   &   \cr
    & \ m^2 = 0.01   && 0.00037  &  0.00037   &   \cr
     \noalign{\hrule}
                                                             }}}
\smallskip
{\bf Table 1:}  Values of $\mu$ for a few representative values of $m^2$ 
and $\phi_0$.
\endinsert

Several representative values of $\mu$ are tabulated in Table 1.
We see that none of these cases yields $\mu > 1$, needed for the
kinematic argument for forming naked singularities. The table
suggests that the most promising region to consider is that of
large masses and large fields\foot{The value of $\po$ is bounded
from above by the requirement that $V(\po) < 0$, which effectively
restricts us to the small field regime for large $m^2$.}. However,
closer exploration of this region did not yield any parameters
giving rise to $\mu > 1$.

\appendix{C}{Probe brane potentials}

We are interested in the supergravity dual to $\CN =4$ SYM
deformed by the operator $\CO_2$. Our strategy will be to start
with  five dimensional gauged supergravity and lift the solution
to  ten dimensional Type IIB supergravity. By the standard AdS/CFT
dictionary, we need to find a solution to the scalar-gravity
system \HHMaction, with the scalar filed profile asymptotically
behaving as the non-normalizable mode.

The deformation by $\CO_2$ will result in a static, spherically
symmetric configuration. So we can use a metric ansatz identical
to that of the scalar hair black hole \scalarhbh\ in Appendix A,
\ie,
\eqn\fived{ ds_5^2 = - e^{-2 \delta(r)} \, f(r) \, dt^2 + {1\over
f(r) } \, dr^2 + r^2 \, d \Omega_3^2 }
We will also have a scalar field profile $\phi(r)$. The equations
of motion are as in \scbheqns.

Given this solution it is simple to write down the full
10-dimensional geometry using the consistent truncation ansatz.
Let $\xi$ be a coordinate on the $\S^5$, so that the round $\S^5$
metric takes the form:
\eqn\sf{ d\Omega_5^2 = d\xi^2 + \sin^2  \xi \, d\psi^2 + \cos^2
\xi \, d\Omega_3^2 \ . }
The full 10-d solution is read off from \clpst\ to be
\eqn\tendm{ ds_{10}^2 = \Delta \, ds_5^2 + F \, \Delta \, d\xi^2 +
{F^2 \over \Delta} \, \sin^2 \, \xi \, d\psi^2 + {1 \over F
\Delta} \, \cos^2 \xi \, d\Omega_3^2 }
where
\eqn\defns{ \Delta^2 = {1\over F^2} \sin^2 \xi + F \cos^2 \xi \ ,
\qquad
 \qquad F = \exp\({1\over 2 \sqrt{3}}\, \phi\)}
The five-form is given as
\eqn\tendff{ G_5 = - U \, \epsilon_5 - 3 \, \sin \xi \, \cos \xi
\; {1\over F} * dF \wedge d\xi \ , }
with
\eqn\udef{ U = - 2 \, \( F^2 \, \cos^2 \xi + {1\over F} \( 1+
\sin^2 \xi \) \)
 \ .}
The $\epsilon_5$ and the Hodge dual are with respect to the five
dimensional metric $ds_5^2$. In particular,
\eqn\epsilonf{ \epsilon_5 = - r^3 \, e^{-\delta}  \, dt \wedge dr
\wedge d\Omega_3 \ . }

The brane probe action is composed of the DBI term and the WZ term
as:
\eqn\braneprobe{ S_{D3} = - \tau_3 \int \,d^4 \zeta \,\sqrt{{\rm
det} \, G^{(ind)}} + \mu_3 \int \, C_4}
We  use static gauge $\zeta^0 = t$, $\zeta^i = \Omega_3^i$ and
evaluate
\eqn\dbi{ \sqrt{{\rm det}\,  G^{(ind)}} = r^3 \,\sqrt{f} \,
\Delta^2 \, e^{-\delta} }
To evaluate the four form $C_4 = C(r,\xi) \, dt \wedge d \Omega_3$
we use
\eqn\Cdef{ {\partial C(r,\xi) \over \partial r} = - \, U \, r^3 \,
e^{-\delta} }

To find the exact probe potential we need to know the precise form
of the functions $F(r)$, $f(r)$ and $\delta(r)$. This can be done
numerically, but it is relatively simple to extract the asymptotic
form of the potential for large $r$. To this end, realize that our
asymptotic boundary conditions are:
\eqn\asymbc{ \eqalign{ f(r) &\sim 1 + r^2  \cr \phi(r) &\sim
\alpha \, {\log(r) \over r^2} \cr \delta(r) &\sim  {1 \over 3} \,
\alpha^2 \, {\log(r)^2 \over r^4} }}
The behaviour of $\delta(r)$ is determined by its equation of
motion \scbheqns. This will not be essential in what follows as
its effect is at a subleading order.

With these ingredients in hand, one simply expands out the
contributions from the DBI term \dbi\ and the WZ term (which is
just $C(r,\xi)$) to the potential. We have
\eqn\dbicont{ r^3 \,\sqrt{f} \,  \Delta^2 \, e^{-\delta} \sim r^4
+ {\alpha \over \sqrt{3} } \, r^2 \, \log(r) \, \( {1\over 2}
\cos^2 \xi - \sin^2 \xi \)+ {1 \over 2} \,r^2 + \cdots }
\eqn\wzcont{ C(r,\xi) = r^4 + {\alpha \over \sqrt{3}} \, \( r^2 \,
\log(r) - {1\over 2} r^2 \) \, \(\cos^2 \xi - {1 \over 2} \(
\sin^2 \xi +1 \) \) + \cdots }
Putting both of these together and using the fact that $\mu_3 =
\tau_3$ for a BPS D3-brane, we see that  the leading order terms
cancel among the two contributions. This is guaranteed by the
supersymmetries preserved by the AdS background, since at this
order we are not picking up any contributions from the deformation
$\phi(r)$. As a result we we are left with an effective potential:
\eqn\effpot{ V_{eff} = {\tau_3 \over 2} \, r^2 \, \( 1 - {\alpha
\over 2 \sqrt{3} } \, \( 3 \, \sin^2 \xi  -1 \) \) }
We see that for $\a > \a_c = 2 \sqrt{3}$, the potential is
unbounded from below. A similar computation done in the case of
the Poincare patch of AdS by \bce\ reveals that the critical value
of $\a$ for the field theory on $\R^4$ is zero, \ie, the field
theory is always unstable under the deformation. Note that the
salubrious term ${\tau_3 \over 2}  \, r^2$ arises here because of
the curvature of the $\S^3$ and can be traced to the constant in
$f(r) = 1 +r^2$.

\listrefs

\end